\documentclass[twocolums,showpacs,showkeys,prb]{revtex4}
\usepackage{amssymb}
\usepackage{amsmath}
\usepackage{graphicx}
\usepackage[normalem]{ulem}
\usepackage{setspace}

\begin{document}

\title{Mesoscopic non-equilibrium thermodynamics approach to non-Debye dielectric relaxation}

\author{Humberto H\'{\i}jar\footnote{Fellow of SNI Mexico. 
E-mail for correspondence: hijar@daad-alumni.de}}
\affiliation{Facultad de Ciencias, Universidad Nacional Aut\'onoma de M\'exico, Circuito
Exterior de Ciudad Universitaria, 04510, D. F., M\'exico}
\author{J. G. M\'endez-Berm\'udez }
\affiliation{ Instituto de F\'{\i}sica, B. Universidad Aut\'onoma de Puebla.
Av. San Claudio y 18 Sur, 72570, Puebla, M\'exico}
\author{Iv\'{a}n Santamar\'{\i}a-Holek\footnote{Fellow of SNI Mexico.
E-mail for correspondence: isholek.fc@gmail.com}}
\affiliation{Facultad de Ciencias, Universidad Nacional Aut\'onoma de M\'exico, Circuito
Exterior de Ciudad Universitaria, 04510, D. F., M\'exico}

\begin{abstract}
Mesoscopic non-equilibrium thermodynamics is used to formulate a model
describing non-homogeneous and non-Debye dielectric relaxation. The model is presented in terms 
of a Fokker-Planck equation for the  probability distribution of non-interacting polar molecules in 
contact with a heat bath and in the presence of an external time-dependent electric field. 
Memory effects are introduced in the Fokker-Planck description through integral relations containing 
memory kernels, which in turn are used to establish a connection with fractional Fokker-Planck 
descriptions. The model is developed in terms of the evolution equations for the first 
two moments of the distribution function. These equations are solved by following a perturbative method
from which the expressions for the complex susceptibilities are obtained as a functions of the
frequency and the wave number. Different memory kernels are considered and used to compare with  
experiments of dielectric relaxation in glassy systems. For the case of Cole-Cole relaxation, 
we infer the distribution of relaxation times and its relation with an effective distribution of 
dipolar moments that can be attributed to different segmental motions of the polymer chains in a melt.
\end{abstract}
\maketitle

\section{Introduction}

Relaxation processes in complex systems such as glass-forming liquids, 
polymeric melts, liquid crystals and even lipid membranes is a  very active 
area of research~\cite{funke,kumar,kob,yin1,hernandez,macro,dejardin001,dejardin002,%
Adam99,oliveira,hershkovits001}. They have been studied using 
different techniques ranging from NMR and 
Brillouin light scattering to dielectric spectroscopy~\cite{yin1,macro}. 
The variety of relaxation phenomena occurring in these systems are non-Debye 
and therefore cannot be characterized by a single relaxation 
time~\cite{Adam99,oliveira}. These multiscale time relaxations
suggest an accumulative effect of different elemental relaxation processes 
that can be associated with collective motions in the 
system~\cite{macro,Adam99,oliveira,hernandez,hershkovits001}.

The average effects of these cooperative mechanisms observed in experiments can be 
accurately described by means of well known empirical relations such as the 
Williams-Watts function in time space, or through its conjugated frequency 
domain (Havrilak-Nagami) function. These functions, and many others that can be found in the 
literature~\cite{colmenero,funke,kumar,kob,montroll,jonscher001}, reflect the specific features of
the micro- or mesoscopic relaxation dynamics of each system. The microscopic description of these 
features may involve the formulation of specific molecule-molecule interaction models that lead
to particular results. In view of this, it seems important to have a general 
theoretical model able to cope with different relaxation processes that may 
describe, for example, the dielectric relaxation in diverse materials including 
solids, glass-forming liquids and polymer melts~\cite{jonscher001}. These last systems 
may exhibit, for example, anomalous Kerr effect which can be described in terms of 
models based on fractional Fokker-Planck equations 
(FFPE)~\cite{dejardin001,dejardin002}.

In this work we adopt a mesoscopic point of view for describing the
relaxation of a system of non-interacting polar molecules inmersed in a
time-dependent external electric field, because at this level of 
description it is still possible to formulate the mentioned general 
framework able to cope with different relaxation dynamics. In particular, 
we use the mesoscopic non-equilibrium thermodynamics (MNET) formalism in 
order to derive a non-Markovian Fokker-Planck like equation (NMFPE) for 
the evolution of the probability distribution describing the orientation 
and position of the molecules. The MNET method uses the entropy produced 
by the system when it is driven out from equilibrium and is particularly 
useful for describing the stochastic evolution of systems at the 
mesoscopic scale~\cite{revMNET,ivanjcp,agustinjcp,mnetPolim}. The advantage of this
formalism lies in the fact that the NMFPE derived has a kinetic origin, and therefore there is no
explicit assumption on the specific time correlation properties of the underlying Gaussian stochastic process,
contrary to what happens with equations derived from Langevin formalisms \cite{coffeyLibro,Talkner}.
Within this framework, linear relations between 
generalized forces and fluxes occurring in phase space are postulated, which 
allow us to introduce memory effects accounting for the non-instantaneous 
response of the system to the external perturbations. The effects of the 
non-Markovian dynamics on a measurable property of the system like its complex 
dielectric susceptibility can then be analyzed in a systematic way by considering 
different memory kernels. In this work we will concentrate mainly on memory kernels 
decaying as power laws. We will show that these particular functions turn out to 
be important since they lead to the mentioned FFPE.

In the section~II we will derive the NMFPE for a system of non-interacting
polar molecules by using MNET. This equation describes both  rotational diffusion and 
the diffusion of polarization, and we will show that it can be recast into a FFPE 
when proper memory kernels are considered. In section~III we will construct 
the equations for the first two moments of distribution. Subsequently, in 
Section~IV we will present a perturbation procedure which allow for a solution of the obtained 
dynamic equations, when the system is subjected to an external harmonic 
electric field superposed to a large static electric field. Consequently, we 
will obtain a closed expression for the non-homogeneous dielectric susceptibility 
of the system which incorporates non-Markovian effects, and the effects of the 
three first moments of the distribution function. These effects will be analyzed 
systematically in section~V by considering diverse limiting cases and memory functions. 
Memory kernels decaying as power laws in time will be shown to generate FFPE and 
response functions of the Cole-Cole type~\cite{cole001}. A comparison with recent
experimental results is performed to validate the models derived. In section~VI we will 
propose a model based on a theory of distribution of relaxation times which can  
be used to give an insight of the physical processes producing power law kernels 
and, accordingly, FFPE. Finally, we will present some final remarks an discuss 
the limitations of our model in section~VII.

\section{Non-Markovian Description of a System of Polar Molecules}

We shall consider a system of $N$ non-interacting polar molecules with
dielectric moment, $\vec{p}$, of magnitude $p$, which are in contact with 
a thermal bath at temperature $T$. For convenience, in the following we will 
use the unit vector, $\vec{n}=\vec{p}/p$, which represents the orientation of 
the molecular dielectric moment, and describe the system in terms of a 
single-molecule probability density. Let 
$f\left( \vec{r},\vec{n},t\right) d\vec{r}d\vec{n}$ denote the probability 
for finding one molecule at a position ranging from $\vec{r}$ to 
$\vec{r}+d\vec{r}$, and oriented within the range $\vec{n}$ and 
$\vec{n}+d\vec{n}$, at time $t$.

The normalization condition,
\begin{equation}
\iint d\vec{r}d\vec{n}\ f\left( \vec{r},\vec{n},t\right) 
=\mathcal{N}\text{, for all }t\text{,}  
\label{mnet000}
\end{equation}%
where $\mathcal{N}$ is the numerical density of molecules, imposes the
following restriction on the time evolution of the probability density
function $f$, \cite{kroeger001,krogerbook}
\begin{equation}
\frac{\partial f}{\partial t}=-\nabla _{i}(fV_{i})-\hat{R}_{i}\left( f\Omega
_{i}\right) .  \label{mnet001}
\end{equation}%
Here, $V_{i}$ and $\Omega _{i}$ are the conjugate velocities to $x_{i}$ and $%
n_{i}$, respectively; $\nabla _{i}$ is the $i$-th spatial derivative; and $%
\hat{R}_{i}$ is the $i$-th component of the so called rotational operator%
\begin{equation*}
\hat{R}_{i}=\varepsilon _{ijk}n_{j}\frac{\partial }{\partial n_{k}}.
\end{equation*}%
In this expression $\varepsilon _{ijk}$ represents the Levi-Civitta
antisymmetric symbol and summation over repeated indices will be implicit
in the next.

Consider now, that an external time-dependent and uniform electric field, 
$\vec{E}\left(t\right)$, acts on the the system. Then, the potential energy 
of one molecule in the presence of this
field, $U\left( \vec{n};t\right) $, is an explicit function of the
orientation vector and also depends on time, $t$, since the
electric field, $\vec{E}\left( t\right) $, is time-dependent. 
In first approximation, the 
potential energy of the system is given by
\begin{equation}\label{U(t)}
U\left( \vec{n};t\right) =-p\vec{n}\cdot \vec{E}\left( t\right) ,
\end{equation}
which holds for molecules in which  the anisotropy of the electric 
polarizability is negligible~\cite{doi001}.

The explicit form of Eq. (\ref{mnet001}) in terms of $f\left( \vec{r},\vec{n},t\right)$ 
can be deduced by analyzing the entropy produced by the system during its time evolution. 
According to the so called Gibbs Entropy Postulate~\cite{revMNET,ivanjcp}, entropy is defined in
terms of the probability distribution $f$ by
\begin{equation}
S=-k_{B}\iint d\vec{r}d\vec{n}\ f\left( \vec{r},\vec{n},t\right) \ln \left[ 
\frac{f\left( \vec{r},\vec{n},t\right) }{f^{\text{leq}}\left( \vec{n}%
;t\right) }\right] +S^{\text{leq}},  \label{mnet003}
\end{equation}%
where $k_B$ is the Boltzmann constant and $S^{\text{leq}}$ is the local-equilibrium entropy. 
According to Eq. (\ref{U(t)}), the corresponding probability density in local-equilibrium is given by
canonical distribution function 
\begin{equation}
f^{\text{leq}}\left( \vec{n};t\right) =\frac{1}{z^{\text{leq}}\left(
T;t\right) }\exp \left[ -\frac{U\left( \vec{n};t\right) }{k_{B}T}\right] ,
\label{mnet002}
\end{equation}%
where $z^{\text{leq}}$ is the corresponding partition function.

In order to find the entropy production, $\sigma =\partial \left( S-S^{\text{leq}}\right)
/\partial t$, we take the time derivative of Eq. (\ref{mnet002}) from where, by 
using the normalization condition (\ref{mnet000}) and Eq.~(\ref{mnet001}), we find 
\begin{equation}
\sigma =\frac{1}{T}\iint d\vec{r}d\vec{n}\ \left[ \mu \nabla _{i}\left( f\
V_{i}\right) +\mu \hat{R}_{i}\left( f\ \Omega _{i}\right) \right]
+ k_{B}\iint d\vec{r}d\vec{n}\ f\ \frac{\partial }{\partial t}
  \ln f^{\text{leq}},  \label{mnet004}
\end{equation}
where the non-equilibrium chemical potential, $\mu =\mu \left( \vec{r},\vec{n},t\right) $,
has been defined by
\begin{equation}
\mu \left( \vec{r},\vec{n},t\right) \equiv k_{B}T\ln \left[ \frac{f\left( \vec{r},%
\vec{n},t\right) }{f^{\text{leq}}\left( \vec{n};t\right) }\right] .
\label{mnet005}
\end{equation}

Now, in accordance with the assumption of local equilibrium, we will consider external 
fields that evolve in a time-scale larger than the time-scale of fluctuations. 
Therefore, changes of the local equilibrium distribution are not significant over the 
short time-scale of the fluctuations, and it can be approximated by
$f^{\text{leq}}\left( \vec{n};t\right)\simeq f^{\text{leq}}\left( \vec{n};0\right)$.
Consequently, the second term on the right hand side of Eq.~(\ref{mnet004}) 
vanishes. This approximation is also supported by the fact that 
later on we will consider electric fields consisting of the superposition
of a static, $\vec{E}^{(0)}$, and a time varying, $\vec{E}^{(1)}$, contributions
such that the magnitude of the time-dependent superposed field 
$\vec{\mathcal{E}}^{(1)}(t)$ is small when compared with the static field 
$\vec{E}^{(0)}$, that is, we will assume 
$\vec{E}(t)=\vec{E}^{(0)}+\vec{\mathcal{E}}^{(1)}(t)$
with $|\vec{\mathcal{E}}^{(1)}|/|\vec{E}^{(0)}|<<1$.

Under this assumption, an integration by parts of Eq.~(\ref{mnet004}) yields the
following expression for $\sigma $, 
\begin{equation}
\sigma =-\frac{1}{T}\iint d\vec{r}d\vec{n}\ \left( fV_{i}\nabla _{i}\mu
+f\Omega _{i}\hat{R}_{i}\mu \right) ,  \label{mnet005a}
\end{equation}%
where we have assumed an infinite system with vanishing probability density
at the boundaries in $\vec{r}$-space, and periodic conditions in $\vec{n}$-space.

The entropy production given by Eq.~(\ref{mnet005a}) involves the sum of products
of the generalized currents, $V_{i}$ and $\Omega _{i}$, and forces, 
$\nabla_{i}\mu $ and $\hat{R}_{i}\mu $. Thus, following a linear response scheme \cite{kubo001}, 
we may assume that the currents are proportional to forces and a linear relation for $V_{i}$ and 
$\Omega _{i}$, in terms of $\nabla _{i}\mu $ and $\hat{R}_{i}\mu $, can be
written. For simplicity, we will not consider cross terms in these relations,
i.e. we will assume that the generalized force $\nabla _{i}\mu $ is the
unique responsible for producing the flux $V_{i}$, and correspondingly for 
$\hat{R}_{i}\mu $ and $\Omega _{i}$. This assumption is in accordance with the 
corresponding Onsager-Casimir reciprocal relations given the vectorial character of the vectors 
involved in the description, see Ref. \cite{gyar}. In the general case and according to the phenomenology
observed, we expect that the response of the system is not instantaneous, and then memory effects
should be considered. Following Refs. \cite{zwan61,jpcb}, these effects can be taken into account by 
introducing memory functions in the following linear relationships
\begin{equation}
fV_{i} = -\int\limits_{-\infty }^{t}ds\ K_{ij}\left( t-s\right) f\left( s
\right) \nabla _{j} \mu\left( s\right) ,  \label{mnet006}
\end{equation}
\begin{equation}
f\Omega _{i}=-\int\limits_{-\infty }^{t}ds\ L_{ij}\left( t-s\right) f\left(
s \right) \hat{R}_{j}\mu \left( s\right) .  \label{mnet007}
\end{equation}

The memory functions $K_{ij}\left( t\right) $ and $L_{ij}\left( t\right) $,
are such that $K_{ij}\left( t\right) =L_{ij}\left( t\right) =0$, for $t<0$.
They describe the delay in the response of the system to forces which vary
in time. This delay is always present and is particularly important when
forces vary fast in time~\cite{kubo001}.

By replacing Eqs.~(\ref{mnet006}) and (\ref{mnet007}) into 
Eq.~(\ref{mnet001}), we obtain 
\begin{equation}
\frac{\partial f}{\partial t}=\nabla _{i}\int_{-\infty }^{t}ds\ \varphi
_{ij}\left( t-s\right) \nabla _{j}f\left( s\right) +\hat{R}_{i}\int_{-\infty
}^{t}ds\ \phi _{ij}\left( t-s\right) \left[ \hat{R}_{j}f\left( s\right) +%
\frac{f\left( s\right) }{k_{B}T}\hat{R}_{j}U\right]  \label{mnet008}
\end{equation}%
where for notation convenience, we have introduced the modified memory functions 
$\varphi _{ij}\left(t\right) =k_{B}TK_{ij}\left( t\right) $, and $\phi _{ij}\left( t\right)
=k_{B}TL_{ij}\left( t\right) $. In order to obtain Eq.~(\ref{mnet008}), we
have also used the relations 
\begin{equation}\label{explicitforces}
f\nabla _{i}\mu =k_{B}T\nabla _{i}f, \hspace*{5mm}\mbox{and}\hspace*{5mm}
f\hat{R}_{i}\mu =k_{B}T\hat{R}_{i}f+f\hat{R}_{i}U,
\end{equation}%
which follow directly from Eq.~(\ref{mnet005}).

Eq.~(\ref{mnet008}) is the desired non-Markovian FP equation for the probability
distribution function $f$. It provides a complete probabilistic description
of the system at the mesoscopic level and constitutes a generalization of previous equations
obtained in the literature for systems with Debye relaxation dynamics \cite{kroeger001,mendez001}. 
The memory kernels $\varphi
_{ij}\left( t\right) $ and $\phi _{ij}\left( t\right) $, characterize the
translational and rotational diffusion processes in the system,
respectively. 
For anisotropic molecules such as rigid polymers in solution, the translational
diffusion tensor depends in general on the orientational degrees of freedom 
$n_{i}$~\cite{doi001}. Here, for simplicity, we will restrict ourselves to consider
isotropic translational and rotational diffusion processes, i.e. $\varphi
_{ij}\left( t\right) =\varphi \left( t\right) \delta _{ij}$, and $\phi
_{ij}\left( t\right) =\phi \left( t\right) \delta _{ij}$. Consequently, our
results will be expected to describe better the dynamics of slightly
anisotropic polar molecules. Within this approximation Eq.~(\ref{mnet008})
can be recast in the final form 
\begin{equation}
\frac{\partial f}{\partial t}=\nabla ^{2}\int_{-\infty }^{t}ds\ \varphi
\left( t-s\right) f\left( s\right) +\hat{R}_{i}\int_{-\infty }^{t}ds\ \phi
\left( t-s\right) \left[ \hat{R}_{i}f\left( s\right) +\frac{f\left( s\right) 
}{k_{B}T}\hat{R}_{i}U\right] .  \label{mnet009}
\end{equation}

In addition, notice that we have restricted ourselves to consider memory kernels
which only depend on time, and accordingly, we have neglected the possible
non-local effects from interactions at different points in space.

In the following sections, Eq. (\ref{mnet009}) will be used to obtain a 
dynamical model for the time evolution of the average polarization $P_i$ and 
its associated order parameter, the quadrupole moment $S_{ij}$.

\subsection{Homogeneous Approximation and Fractional Fokker-Planck Equation}

Diverse interesting limiting cases can be obtained from Eq.~(\ref{mnet009}).
Firstly, we will consider the so called homogeneous approximation~\cite%
{mendez001}, in which a simplified description in terms of the reduced
orientational probability distribution, 
\begin{equation*}
g\left( \vec{n},t\right) =\int d\vec{r}\ f\left( \vec{r},\vec{n},t\right) ,
\end{equation*}%
is given. Averaging Eq.~(\ref{mnet009}) over space coordinates, the corresponding
non-Markovian FP equation for $g$ is found to be 
\begin{equation}
\frac{\partial g}{\partial t}=\hat{R}_{i}\int_{-\infty }^{t}ds\ \phi \left(
t-s\right) \left[ \hat{R}_{i}g\left( s\right) +\frac{g\left( s\right) }{%
k_{B}T}\hat{R}_{i}U\right] .  \label{frac001}
\end{equation}

Clearly, moments of the reduced distribution $g$ will be uniform
time-dependent quantities. Therefore, in the homogeneous approximation spatial diffusion of
these moments is ignored and the properties obtained from Eq.~(\ref{frac001})
must be considered as volume averages~\cite{mendez001}.

It is interesting to notice that Eq.~(\ref{frac001}) leads to a FFPE when the 
memory function $\phi \left( t\right) $ is conveniently chosen to be a power 
law. Specifically, let
\begin{equation}
\phi \left( t\right) =\left\{ 
\begin{array}{cc}
\frac{D^{2}}{\Gamma \left( \alpha -1\right) }\left( Dt\right) ^{\alpha -2} & 
\text{for }t>0 \\ 
0 & \text{for }t<0%
\end{array}%
\right. ,  \label{frac002}
\end{equation}%
where $\Gamma $ represents the gamma function, $\alpha $ is a dimensionless
parameter within the range $0<\alpha <1$, and $D$ is a
constant with units of inverse time that can be identified with the
rotational diffusion coefficient of the molecules, as we will show below. 
It follows directly from Eq.~(\ref{frac001}) that when the memory function 
is given by Eq.~(\ref{frac002}), the orientational distribution obeys the 
dynamic equation 
\begin{equation}
\frac{\partial g}{\partial t}=\frac{D^{\alpha }}{\Gamma\left( \alpha
-1\right)} \hat{R}_{i} \int\limits_{-\infty}^{t} ds\, \frac{1}{\left( t -
s\right)^{2-\alpha}} \left[ \hat{R}_{i}g\left( s \right) + \frac{g\left( s
\right) }{k_{B}T}\hat{R}_{i}U\right] .  \label{frac002a}
\end{equation}

On the right hand side of the last expression we can identify the
generalized fractional derivative operator \cite{dejardin001,klafter}
\begin{equation*}
{}_{-\infty}\!D_{t}^{ -\nu } g\left( t\right) = \frac{1}{\Gamma\left(
\nu\right)} \int\limits_{-\infty}^{t}ds\, \frac{g\left( s\right)}{\left( t -
s\right)^{1-\nu}} , 
\end{equation*}
for $\nu = \alpha -1$. Thus Eq.~(\ref{frac002a}) can be recast in the form 
\begin{equation}
\frac{\partial g}{\partial t}=D^{\alpha } {}_{-\infty}\!D_{t}^{1-\alpha} 
\hat{R}_{i}\left[ \hat{R}_{i}g+\frac{g}{k_{B}T}\hat{R}_{i}U\right] ,
\label{frac006}
\end{equation}
which is the referred FFPE for the orientational distribution.

It can be noticed that Eq.~(\ref{frac006}) reduces to the usual Fokker-Planck 
equation for $\alpha =1$ (normal rotational diffusion),
\begin{equation*}
\frac{\partial g}{\partial t}= D\hat{R}_{i}
                             \left[ 
                             \hat{R}_{i}g+\frac{g}{k_{B}T}\hat{R}_{i}U
                             \right] ,
\end{equation*}
where $D$ can be identified as the rotational diffusion 
coefficient~\cite{doi001}. For $0<\alpha <1$, Eq.~(\ref{frac006}) describes 
the non-Markovian behavior associated with rotational subdiffusive processes.

An interesting point concerning the applicability of the memory 
function~(\ref{frac002}) is worth stressing. The
presence of $\Gamma \left( \alpha -1\right) $ on the right hand side of Eq.~(%
\ref{frac002}) yields a negative memory function for $0<\alpha <1$. This
implies that the memory function given by Eq.~(\ref{frac002}) superposes all
the effects prior time $t$ with an intrinsic negative sign. This fact is
rather masked by the fractional derivative operator appearing in Eq.~(\ref%
{frac006}), and as far as we know it has not been noticed in literature
where dynamic equations of the type~(\ref{frac006}) are postulated for the
probability distribution $g$. The presence of
the function $\Gamma \left( \alpha -1\right) $ is, however, necessary in
order to obtain the FFPE, Eq.~(\ref{frac002a}), from a power law
decay of the memory function.

Finally, we will consider the case in which both rotational and translational
diffusion processes are present and can be described by power law decaying
memory functions. For the translational kernel we introduce the function
\begin{equation}
\varphi \left( t\right) =\left\{ 
\begin{array}{cc}
\frac{\mathcal{D}_{\beta }}{\Gamma \left( \beta -1\right) }t^{\beta -2} & 
\text{for }t>0 \\ 
0 & \text{for }t<0%
\end{array}%
\right. ,  \label{frac007}
\end{equation}
where $0<\beta <1$, is a constant parameter, and $\mathcal{D}_{\beta }$ is a
generalized translational diffusion coefficient with units of length square
times $t^{-\beta }$\cite{metzler001}.

By substituting Eqs.~(\ref{frac002}) and (\ref{frac007}) into
Eq.~(\ref{mnet009}), it can be easily shown that the FPE describing the
evolution of the probability density $f$ takes the form of a fractional
equation,
\begin{equation}
\frac{\partial f}{\partial t}=\!\mathcal{D}_{\beta }{}_{-\infty
}\!D_{t}^{1-\beta }\nabla ^{2}f+D^{\alpha }{}_{-\infty }\!D_{t}^{1-\alpha }%
\hat{R}_{i}\left[ \hat{R}_{i}f+\frac{f}{k_{B}T}\hat{R}_{i}U\right] ,
\label{frac008}
\end{equation}%
which includes anomalous diffusion through the fractional derivative
${}_{-\infty}D_{t}^{1-\beta }$. In the limit $\alpha ,\beta =1$, the Markovian
approximation is recovered.

\section{Dynamic Equations for the Multipolar Moments}

At the macroscopic level, the description of the system must be given in
terms of the moments of the distribution function $f$. For this purpose, 
Eq.~(\ref{mnet008}) can be used to construct a 
hierarchy of equations for the multipolar moments associated with the charge
distribution. Let us consider the dipole, $P_{i}$, quadrupole, $S_{ij}$, and
octupole, $Q_{ijk}$ fields, which are defined by the following averages
\begin{equation}
P_{i}\left( \vec{r},t\right) =\mathcal{N}\int d\vec{n}\,f\left( \vec{r},\vec{%
n},t\right) p_{i},  \label{multi001}
\end{equation}%
\begin{equation}
S_{ij}\left( \vec{r},t\right) =\frac{1}{2}\mathcal{N}^{2}\int d\vec{n}%
\,f\left( \vec{r},\vec{n},t\right) \left( 3p_{i}p_{j}-p^{2}\delta
_{ij}\right) ,  \label{multi002}
\end{equation}%
\begin{equation}
Q_{ijk}\left( \vec{r},t\right) =\frac{1}{2}\mathcal{N}^{3}\int d\vec{n}%
\,f\left( \vec{r},\vec{n},t\right) \left[ 5p_{i}p_{j}p_{k}-3p^{2}\left\{ 
\mathbf{1}\vec{p}\right\} _{ijk}\right] .  \label{multi003}
\end{equation}

In the last definition, $\left\{ \mathbf{1}\vec{p}\right\} $ represents the
symmetric part of the tensor $\mathbf{1}\vec{p}$, which has the explicit form
\begin{equation*}
\left\{ \mathbf{1}\vec{p}\right\} _{ijk}=\frac{1}{3}\left( p_{i}\delta
_{jk}+p_{j}\delta _{ik}+p_{k}\delta _{ij}\right) .
\end{equation*}

Multiplying Eq.~(\ref{mnet008}) by $\mathcal{N}p_{i}$ and averaging
over the orientational degrees of freedom we obtain
\begin{eqnarray}
\frac{\partial P_{i}}{\partial t} &=&\nabla
^{2}\int\limits_{-\infty}^{t}ds\, \varphi \left( t-s\right) P_{i}\left(
s\right) -2\int\limits_{-\infty }^{t}ds\, \phi \left( t-s\right) 
P_{i}\left( s\right)  \label{multi004} \\
&& +\frac{2\mathcal{N}p^{2}}{3k_{B}T} \int\limits_{-\infty }^{t}ds\,\phi
\left( t-s\right) E_{i}\left( s\right) -\frac{2}{3\mathcal{N}k_{B}T}
\int\limits_{-\infty }^{t}ds\, \phi \left(t-s\right) S_{ij}\left( s\right)
E_{j}\left( s\right) .  \notag
\end{eqnarray}
The first term at the right hand side of the last equation accounts for 
the diffusion of the polarization whereas the other terms are consequences 
of the orientational dynamics.

Analogously, multiplying Eq.~(\ref{mnet008}) by $\mathcal{N}^{2}\left(
3p_{i}p_{j}-p^{2}\delta _{ij}\right) /2$ and averaging over $\hat{n}$, we obtain the
dynamic equation for the quadrupolar moment,%
\begin{eqnarray}
\frac{\partial S_{ij}}{\partial t} &=&\nabla ^{2}\int\limits_{-\infty
}^{t}ds\,\varphi \left( t-s\right) S_{ij}\left( s\right)
-6\int\limits_{-\infty }^{t}ds\,\phi \left( t-s\right) S_{ij}\left( s\right) 
\label{multi005} \\
&&+\frac{9\mathcal{N}p^{2}}{10k_{B}T}\int\limits_{-\infty }^{t}ds\,\phi
\left( t-s\right) \left[ \!\left[ \vec{P}\left( s\right) \vec{E}\left(
s\right) \right] \!\right] _{ij}-\frac{6}{5\mathcal{N}k_{B}T}%
\int\limits_{-\infty }^{t}ds\,\phi \left( t-s\right) Q_{ijk}\left( s\right)
E_{k}\left( s\right) .  \notag
\end{eqnarray}%
where $\left[ \!\left[ \vec{P}\left( t\right) \vec{E}\left( t\right) \right]
\!\right] $ is the symmetric traceless part of the dyad $\vec{P}\vec{E}$
evaluated at time $t$, i.e.%
\begin{equation*}
\left[ \!\left[ \vec{P}\left( t\right) \vec{E}\left( t\right) \right] \!%
\right] _{ij}=E_{i}\left( t\right) P_{j}\left( t\right) +E_{j}\left(
t\right) P_{i}\left( t\right) -\frac{2}{3}E_{k}\left( t\right) P_{k}\left(
t\right) \delta _{ij}.
\end{equation*}

It can be noticed from Eq.~(\ref{multi003}) that the contraction 
$Q_{ijk}\left( t\right) E_{k}\left( t\right) $ yields a symmetric traceless
tensor of rank two. Consequently, Eq.~(\ref{multi005}) is in agreement with 
the symmetry properties of the order parameter tensor $S_{ij}$. For 
notation's simplicity, in Eqs.~(\ref{multi004}) and~(\ref{multi005}) we have not
written the dependence of the fields $P_{i}$, $S_{ij}$ and $Q_{ijk}$ on $\vec{r}$.

Notice that Eqs.~(\ref{multi004}) and~(\ref{multi005}) constitute the
first two equations of an infinite recurrence hierarchy which includes 
the time evolution equation for the octupole moment, $Q_{ijk}$, and higher 
order moments. The problem of approximating this infinite hierarchy by a
finite number of equations is a common feature of diverse systems in
non-equilibrium statistical mechanics~\cite{risken001,zwanzigbook}. In 
the following section we will describe in detail our closure approximation 
for the hierarchy of recurrence equations for $P_{i}$, $S_{ij}$, $Q_{ijk}$.

In Eqs.~(\ref{multi004}) and~(\ref{multi005}), the moments $P_{i}$ and 
$S_{ij}$ turn out to be coupled only with moments of
next superior order because translational diffusion has been assumed to be
isotropic. Otherwise the evolution equation for $P_{i}$ will not only
involve $S_{ij}$ but $Q_{ijk}$ as well, while the dynamic equation for $%
S_{ij}$ will include contributions of $Q_{ijk}$ and the fourth
order moment. These contributions have been shown to be important in recent
literature for describing the dynamics of anisotropic molecules like those
composing liquid crystals materials~\cite{kroeger001}.

Eqs.~(\ref{multi004}) and~(\ref{multi005}) can be reduced to the
homogeneous case in which both $P_{i}$ and $S_{ij}$ can be approximated 
by their position-independent volume averages
\begin{equation}
P_{i}^{\text{h}}\left( t\right) =\mathcal{N}\int d\vec{n}\,
                            g\left( \vec{n},t\right) p_{i},  
\label{multi007}
\end{equation}
and 
\begin{equation}
S_{ij}^{\text{h}}\left( t\right) =\frac{1}{2}\mathcal{N}^{2}\int d\vec{n}%
\,g\left( \vec{n},t\right) \left( 3p_{i}p_{j}-p^{2}\delta _{ij}\right) .
\label{multi008}
\end{equation}
Then, it can be proved that $P_{i}^{\text{h}}$ and $S_{ij}^{\text{h}}$ obey 
the dynamical equations
\begin{eqnarray}
\frac{\partial P_{i}^{\text{h}}}{\partial t} &=&-2\int\limits_{-\infty
}^{t}ds\,\phi \left( t-s\right) P_{i}^{\text{h}}\left( s\right) +\frac{2%
\mathcal{N}p^{2}}{3k_{B}T}\int\limits_{-\infty }^{t}ds\,\phi \left(
t-s\right) E_{i}\left( s\right)  \notag \\
&&-\frac{2}{3\mathcal{N}k_{B}T}\int\limits_{-\infty }^{t}ds\,\phi \left(
t-s\right) S_{ij}^{\text{h}}\left( s\right) E_{j}\left( s\right) ,
\label{multi009}
\end{eqnarray}
and 
\begin{eqnarray}
\frac{\partial S_{ij}^{\text{h}}}{\partial t} &=&-6\int\limits_{-%
\infty}^{t}ds\, \phi \left( t-s\right) S_{ij}^{\text{h}}\left( s\right) +%
\frac{9\mathcal{N}p^{2}}{10k_{B}T} \int\limits_{-\infty }^{t}ds\, \phi
\left(t-s\right) \left[ \! \left[ \vec{P}^{\text{h}}\left( s\right) \vec{E}%
\left(s\right) \right] \! \right]_{ij}  \notag \\
&&-\frac{6}{5\mathcal{N}k_{B}T} \int\limits_{-\infty }^{t}ds\, \phi
\left(t-s\right) Q_{ijk}\left( s\right) E_{k}\left( s\right) ,
\label{multi010}
\end{eqnarray}
respectively. Eqs.~(\ref{multi009}) and~(\ref{multi010}) can also be obtained
from Eq.~(\ref{frac001}) for the reduced orientational probability
distribution after multiplying and averaging over $\hat{n}$, as it was done 
in the case of Eqs.~(\ref{multi004}) and~(\ref{multi005}).

\section{Perturbation Theory and Dielectric relaxation}

In the following, we will focus on calculating the response function to a 
small oscillating field 
$\vec{E}^{(1)}\left( t\right) =\mathcal{\vec{E}}^{(1)}e^{i\omega t}$, which 
is superposed to a large static field $\vec{E}^{(0)}$, such that 
$\mathcal{E}^{(1)}/E^{(0)}\ll 1$. Accordingly, the total field acting on the 
system is 
$\vec{E}\left( t\right) =\vec{E}^{(0)}+\mathcal{\vec{E}}^{(1)}e^{i\omega t}$.
In this section we shall show that a perturbation theory in terms of the
small parameter, $\mathcal{E}^{(1)}/E^{(0)}$, can be used in order to find
the frequency and field-dependent dielectric susceptibility of the system. 

With this purpose, let $P_{i}^{(0)}$, $S_{ij}^{(0)}$, etc., denote the
electric multipolar moments associated with the static field. It is clear
that these quantities relax towards equilibrium by following equations which
are completely analogous to Eqs.~(\ref{multi004}) and (\ref{multi005}), with 
$P_{i}^{(0)}$, $S_{ij}^{(0)}$, $Q_{ijk}^{(0)}$ and $E_{i}^{(0)}$, replacing
the fields $P_{i}$, $S_{ij}$, $Q_{ijk}$ and $E_{i}$, respectively.
When both $\vec{E}^{(0)}$ and $\vec{E}^{(1)}\left( t\right) $ act on the system, 
the response to the oscillating field can be identified from the dynamics of the 
perturbed fields $P_{i}^{(1)}=P_{i}-P_{i}^{(0)}$, 
$S_{ij}^{(1)}=S_{ij}-S_{ij}^{(0)}$, etc.~\cite{kubo001}. By substracting the 
equations corresponding to the unperturbed fields from Eqs.~(\ref{multi004}) and 
(\ref{multi005}), the following expressions are found for the time evolution of the 
perturbed fields 
\begin{eqnarray}
\frac{\partial P_{i}^{(1)}}{\partial t} &=&\nabla ^{2}\int\limits_{-\infty
}^{t}ds\,\varphi \left( t-s\right) P_{i}^{(1)}\left( s\right)
-2\int\limits_{-\infty }^{t}ds\,\phi \left( t-s\right) P_{i}^{(1)}\left(
s\right)   \label{perturbation003} \\
&&+\frac{2\mathcal{N}p^{2}}{3k_{B}T}\int\limits_{-\infty }^{t}ds\,\phi
\left( t-s\right) E_{i}^{(1)}\left( s\right)   \notag \\
&&-\frac{2}{3\mathcal{N}k_{B}T}\int\limits_{-\infty }^{t}ds\,\phi \left(
t-s\right) \left[ S_{ij}\left( s\right) E_{j}\left( s\right)
-S_{ij}^{(0)}\left( s\right) E_{j}^{(0)}\right] ,  \notag
\end{eqnarray}
\begin{eqnarray}
\frac{\partial S_{ij}^{(1)}}{\partial t} &=&\nabla ^{2}\int\limits_{-\infty
}^{t}ds\,\varphi \left( t-s\right) S_{ij}^{(1)}\left( s\right)
-6\int\limits_{-\infty }^{t}ds\,\phi \left( t-s\right) S_{ij}^{(1)}\left(
s\right)   \label{perturbation004} \\
&&+\frac{9\mathcal{N}p^{2}}{10k_{B}T}\int\limits_{-\infty }^{t}ds\,\phi
\left( t-s\right) \left[ \!\left[ \vec{P}\left( s\right) \vec{E}\left(
s\right) -\vec{P}^{(0)}\left( s\right) \vec{E}^{(0)}\right] \!\right] _{ij} 
\notag \\
&&-\frac{6}{5\mathcal{N}k_{B}T}\int\limits_{-\infty }^{t}ds\,\phi \left(
t-s\right) \left[ Q_{ijk}\left( s\right) E_{k}\left( s\right)
-Q_{ijk}^{(0)}\left( s\right) E_{k}^{(0)}\right] .  \notag
\end{eqnarray}

For small perturbative fields, $E^{(1)} \ll E^{(0)}$, deviations with
respect to $P^{(0)}_{i}$, $S^{(0)}_{ij}$, $Q^{(0)}_{ijk}$, etc., are
expected to be also small. In this case, we can approximate 
\begin{equation*}
S_{ij}\left( t\right) E_{j}\left( t\right) - S^{(0)}_{ij}\left( t\right)
E^{(0)}_{j} \simeq S^{(0)}_{ij}\left( t\right) E^{(1)}_{j}\left( t\right) +
S^{(1)}_{ij}\left( t\right) E^{(0)}_{j} , 
\end{equation*}
\begin{equation*}
\left[ \! \left[ \vec{P}\left( t\right) \vec{E}\left( t\right)  
-\vec{P}^{(0)}\left( t\right) \vec{E}^{(0)} \right] \! \right]_{ij} 
\simeq  
\left[
\! \left[ \vec{P}^{(0)}\left( t\right) \vec{E}^{(1)}\left( t\right) \right]
\! \right]_{ij} +\left[ \! \left[ \vec{P}^{(1)}\left( t\right) \vec{E}^{(0)}
\right] \! 
\right]_{ij} ,
\end{equation*}
\begin{equation*}
Q_{ijk}\left( t\right) E_{k}\left( t\right) -Q^{(0)}_{ijk}\left( t\right)
E^{(0)}_{k} 
\simeq  
Q^{(0)}_{ijk}\left( t\right) E^{(1)}_{k}\left( t\right)
+Q^{(1)}_{ijk}\left( t\right) E^{(0)}_{k} . 
\end{equation*}

Moreover, for a large static field we can assume that the non-perturbed
multipoles remain close to their equilibrium time-independent values,
i.e. 
$P^{(0)}_{i}\left( t\right)
\simeq P^{\text{eq}}_{i}\left( \vec{E}^{(0)}\right)
\equiv P^{\text{eq}}_{i}$, 
$S^{(0)}_{ij}\left( t\right)
\simeq S^{\text{eq}}_{ij}\left( \vec{E}^{(0)}\right)
\equiv S^{\text{eq}}_{ij}$, 
$Q^{(0)}_{ijk}\left( t\right)
\simeq Q^{\text{eq}}_{ijk}\left( \vec{E}^{(0)}\right)
\equiv Q^{\text{eq}}_{ijk}$,
where, in order to simplify the succeeding discussion, we 
have omitted to write the dependence of the equilibrium moments
on $\vec{E}^{(0)}$. In the following it should be regarded that these 
quantities are of order zero in the perturbative field.

Finally, in order to close the hierarchy of recurrence equations, we will
assume that moments of order $Q_{ijk}$ and higher are not perturbed at all
by the time dependent field, and consequently 
$Q^{(1)}_{ijk}\left( t\right)\simeq 0$, for all $t$.

All these approximations yield the following closed system of equations for
the time evolution of the perturbed dipole, $P^{(1)}_{i}$, and quadrupole, 
$S^{(1)}_{ij}$, moments 
\begin{eqnarray}
\frac{\partial P^{(1)}_{i}}{\partial t} & = & \nabla^{2}
\int\limits_{-\infty}^{t} ds\, \varphi \left( t-s\right) P^{(1)}_{i} \left(
s\right) -2 \int\limits_{-\infty}^{t} ds\, \phi \left( t-s\right)
P^{(1)}_{i} \left( s\right)  \label{perturbation005} \\
&& + \frac{2\mathcal{N}p^{2}}{3 k_{B} T} \int\limits_{-\infty}^{t} ds\, \phi
\left( t-s\right) E^{(1)}_{i}\left( s\right)  \notag \\
&& -\frac{2}{3\mathcal{N}k_{B}T} \int\limits_{-\infty }^{t}ds\, \phi
\left(t-s\right) \left[ S^{\text{eq}}_{ij}
E^{(1)}_{j}\left( s\right) +S^{(1)}_{ij}\left( s\right) E^{(0)}_{j} \right] ,
\notag
\end{eqnarray}
\begin{eqnarray}
\frac{\partial S^{(1)}_{ij}}{\partial t} &=&
\nabla^{2}\int\limits_{-\infty}^{t}ds\, \varphi \left( t-s\right)
S^{(1)}_{ij}\left( s\right) 
-6\int\limits_{-\infty }^{t}ds\, \phi \left( t-s\right) 
S^{(1)}_{ij}\left( s\right)  \label{perturbation006} \\
&& +\frac{9\mathcal{N}p^{2}}{10k_{B}T} 
\int\limits_{-\infty }^{t}ds\, \phi\left( t-s\right) 
\left\{ 
\left[ \! \left[ \vec{P}^{\text{eq}} \vec{E}^{(1)}\left( s\right) \right] \! \right]_{ij} 
+\left[ \! \left[ \vec{P}^{(1)}\left( s\right) \vec{E}^{(0)} \right] \! \right]_{ij}
\right\}  \notag \\
&& -\frac{6}{5\mathcal{N}k_{B}T} \int\limits_{-\infty }^{t}ds\, \phi
\left(t-s\right) Q^{\text{eq}}_{ijk}
E^{(1)}_{k}\left( s\right) .  \notag
\end{eqnarray}

It can be observed that the solution of Eqs.~(\ref{perturbation005}) and 
(\ref{perturbation006}), gives indeed the perturbed polarization as a linear
function of the perturbative field. From this solution the dielectric
susceptibility of the system to the small oscillating field can be
identified. For this purpose, we consider the Fourier transform of 
Eqs.~(\ref{perturbation005}) and (\ref{perturbation006}), which can be 
written in the form 
\begin{equation}
G_{1}\left( \vec{k},\omega \right) P_{i}^{(1)}\left( \vec{k},\omega \right)
= 2\phi \left( \omega \right) 
\left\{ 
\chi _{0}\delta _{ij} -\frac{1}{3\mathcal{N}k_{B}T} 
\left[ S_{ij}^{\text{eq}} E_{j}^{(1)}\left( \vec{k},\omega \right)
     + S_{ij}^{(1)}\left( \vec{k},\omega \right) E_{j}^{(0)} 
\right]
\right\} ,
\label{perturbation007}
\end{equation}
\begin{eqnarray}
G_{2}\left( \vec{k},\omega \right) S_{ij}^{(1)}\left( \vec{k},\omega \right)
&=&\frac{27}{10}\chi _{0}\phi \left( \omega \right)
\left\{
\left[ \!\left[ \vec{P}^{\text{eq}} 
\vec{E}^{(1)}\left( \vec{k},\omega \right) \right] \!\right] _{ij} 
+
\left[ \!\left[ \vec{P}^{(1)}\left( \vec{k},\omega \right)
\vec{E}^{(0)}\right] \!\right] _{ij}
\right\} \notag \\
&&-\frac{6}{5\mathcal{N}k_{B}T}\phi \left( \omega \right) . \label{perturbation008}
\end{eqnarray}

Here we have introduced the notation $\chi _{0}=\mathcal{N}p^{2}/3k_{B}T$,
for the static dielectric susceptibility, and $G_{1}\left( \vec{k},\omega
\right) $ and $G_{2}\left( \vec{k},\omega \right) $ for the propagators in
Fourier space defined by 
\begin{equation}
G_{1}\left( \vec{k},\omega \right) =i\omega +2\phi \left( \omega \right)
+k^{2}\varphi \left( \omega \right) ,  \label{perturbation008a}
\end{equation}%
\begin{equation}
G_{2}\left( \vec{k},\omega \right) =i\omega +6\phi \left( \omega \right)
+k^{2}\varphi \left( \omega \right) .  \label{perturbation008b}
\end{equation}

In Eqs.~(\ref{perturbation007}) and (\ref{perturbation008}) we have also
introduced the notation 
$\vec{E}^{(1)}\left( \vec{k},\omega \right) 
=\left( 2\pi \right) ^{3}\vec{E}^{(1)}\left( \omega \right) 
\delta \left( \vec{k}\right) $, for the space-time Fourier transform of the 
perturbation field. 

The frequency-dependent dielectric susceptibility can be identified by
solving Eq.~(\ref{perturbation008}) for $S_{ij}^{(1)}$, inserting the result
into Eq.~(\ref{perturbation007}), and solving for $P_{i}^{(1)}$. After some
algebraic manipulations this procedure finally yields 
\begin{equation}
P_{i}^{(1)}\left( \vec{k},\omega \right) =\chi _{ij}\left( \vec{k},\omega
\right) E_{j}^{(1)}\left( \vec{k},\omega \right) ,  \label{perturbation010}
\end{equation}%
where the susceptibility, $\chi _{ij}\left( \vec{k},\omega \right) $, can be
written as the product $\chi _{ij}=M_{ik}^{-1}N_{kj}$, with the following
definition for the matrices $M_{ij}$ and $N_{ij}$, 
\begin{equation}
M_{ij}\left( \vec{k},\omega \right) =G_{1}\left( \vec{k},\omega \right)
\delta _{ij}+\frac{p^{2}}{5k_{B}^{2}T^{2}}G_{2}^{-1}\left( \vec{k},\omega
\right) \phi ^{2}\left( \omega \right) \left( 3E^{(0)2}\delta
_{ij}+E_{i}^{(0)}E_{j}^{(0)}\right) , \label{perturbation011}
\end{equation}
\begin{eqnarray}
N_{ij}\left( \vec{k},\omega \right)  
&=&\phi \left( \omega \right) \left\{ 2\chi _{0}\delta _{ij}
-\frac{2}{3\mathcal{N}k_{B}T}S_{ij}^{\text{eq}}
+\frac{4}{5\mathcal{N}^{2}k_{B}^{2}T^{2}}
G_{2}^{-1}\left( \vec{k}, \omega \right) 
\phi \left( \omega \right) Q_{ijk}^{\text{eq}}E_{k}^{(0)}
\right. 
\notag \\
&&
\left.
+\frac{3p^{2}}{5k_{B}^{2}T^{2}}G_{2}^{-1}\left( \vec{k},\omega \right)
\phi \left( \omega \right) \left[ P_{i}^{\text{eq}}E_{j}^{(0)}
+P_{k}^{\text{eq}}E_{k}^{(0)}\delta _{ij}
-\frac{2}{3}P_{j}^{\text{eq}}E_{i}^{(0)}\right]  
\right\} 
.  \label{perturbation012}
\end{eqnarray}

In the following section we will explicitely calculate the complex dielectric
susceptibility, $\chi \left( \vec{k},\omega \right) $, for a specific
geometry of the external fields and explore the effects of the memory
functions, $\phi \left( \omega \right)$ and $\varphi \left( \omega \right)$, 
on this quantity.

\section{Results}

In oder to illustrate the effects of the different mechanisms considered in 
previous sections on the response of the system to the applied fields, in this 
section we consider a specific geometry in which the static electric field, 
$\vec{E}^{(0)}$, is directed along the $\hat{e}_{3}$-axis of a Cartesian 
coordinate system. Furthermore, we will calculate the equilibrium tensors 
$P_{i}^{\text{eq}}$, $S_{ij}^{\text{eq}}$ and $Q_{ijk}^{\text{eq}}$, by using the
stationary solution of the reduced orientational equation, Eq.~(\ref{frac001}),
\begin{equation}
g^{\text{eq}}\left( \vec{n}\right) =
\frac{pE^{(0)}}{4\pi k_{B}T}
\frac{ \exp{\left( {p\vec{n}\cdot \vec{E}^{(0)}}/{k_{B}T}\right) } }
     {\sinh \left( {pE^{(0)}}/{k_{B}T}\right) } 
.  \label{relax001}
\end{equation}

For this probability density and the geometry described
above, we obtain the following expressions for the equilibrium averages
of the moments entering in Eqs.~(\ref{perturbation011}) and 
(\ref{perturbation012}):
\begin{equation}
\left( P_{i}^{\text{eq}}\right) =p\mathcal{N}\left( 
\begin{matrix}
0 \\ 
0 \\ 
L\left( \xi \right) 
\end{matrix}%
\right) ,  \label{illustration001}
\end{equation}%
\begin{equation}
\left( S_{ij}^{\text{eq}}\right) =
p^{2}\mathcal{N}^{2}
\left( 1- \frac{3 L\left( \xi \right)}{\xi} 
\right)
\left( 
\begin{array}{ccc}
-1/2 &   0  & 0 \\ 
0    & -1/2 & 0 \\ 
0    &   0  & 1 
\end{array}
\right) ,  \label{illustration002}
\end{equation}
\begin{equation}
\left( Q_{ijk}^{\text{eq}}E_{k}^{(0)}\right) = p^{3} \mathcal{N}^{3} E^{(0)}%
\left[ \left( 1+\frac{15}{\xi ^{2}}\right) L\left( \xi \right) -\frac{5}{\xi 
}\right] \left( 
\begin{array}{ccc}
-1/2 & 0 & 0 \\ 
0 & -1/2 & 0 \\ 
0 & 0 & 1
\end{array}
\right) .  \label{illustration003}
\end{equation}

In Eqs.~(\ref{illustration001})-(\ref{illustration003}), the dimensionless 
parameter $\xi \equiv pE^{(0)}/k_{B}T$ quantifies the electric energy of a 
molecule in the static field relative to the thermal energy, and 
$L\left( \xi \right) =\coth \left( \xi \right) -1/\xi $ represents the Langevin 
function. As expected, from Eqs.~(\ref{perturbation011}), (\ref{perturbation012})
and~(\ref{illustration001})-(\ref{illustration003}), we obtain that the matrices 
$M_{ij}$, $N_{ij}$ and $\chi_{ij}$ reflect the uniaxial symmetry of the problem 
and are given by
\begin{equation}
\left( M_{ij}\right) = \left( 
\begin{matrix}
M_{\perp} & 0 & 0 \\ 
0 & M_{\perp} & 0 \\ 
0 & 0 & M_{\parallel}%
\end{matrix}
\right) 
, 
\left( N_{ij}\right) = \left( 
\begin{matrix}
N_{\perp} & 0 & 0 \\ 
0 & N_{\perp} & 0 \\ 
0 & 0 & N_{\parallel}%
\end{matrix}
\right)
, \text{ and } 
\left( \chi_{ij}\right) = \left( 
\begin{matrix}
\chi_{\perp} & 0 & 0 \\ 
0 & \chi_{\perp} & 0 \\ 
0 & 0 & \chi_{\parallel}%
\end{matrix}
\right)
.  \nonumber
\end{equation}

After calculating the expressions for the components $M_{\perp}$, $M_{\parallel}$,
$N_{\perp}$ and $N_{\parallel}$ (see the Appendix), the corresponding formulas for 
the components of the dielectric susceptibility perpendicular and parallel to the 
static field are found to be
\begin{equation}
\chi_{\perp }\left( \vec{k},\omega \right) 
=\frac{ \frac{3}{2} \chi_{0}}
{\tilde{G}_{1}\left( \vec{k},\omega \right) 
+ \frac{3}{20}\xi ^{2} \tilde{G}^{-1}_{2}\left( \vec{k},\omega \right)}
\left\{ 1-\frac{L\left( \xi \right) }{\xi } 
+ \tilde{G}^{-1}_{2}\left( \vec{k},\omega \right)
\left[  
1 - \left( \frac{\xi }{2}+\frac{3}{\xi }\right) L\left( \xi \right) 
\right]
\right\} ,
\label{illustration010}
\end{equation}%
\begin{equation}
\chi_{\parallel }\left( \vec{k},\omega \right) 
= \frac{3 \chi_{0} }{ \tilde{G}_{1}\left( \vec{k},\omega \right) 
+ \frac{4}{5}\xi ^{2} \tilde{G}^{-1}_{2}\left( \vec{k},\omega \right) }
\left\{  
\frac{L\left( \xi \right) }{\xi } 
+ \tilde{G}^{-1}_{2}\left( \vec{k},\omega \right) 
\left[ \frac{3}{\xi }L\left( \xi\right) -1  \right] 
\right\} ,  \label{illustration011}
\end{equation} 
where we have introduced the normalized propagators 
$\tilde{G}_{1,2} =G_{1,2}\left( \vec{k},\omega \right) / 2 \phi \left( \omega \right) $. 
The dependence of Eqs.~(\ref{illustration010})-(\ref{illustration011})  on the dimensionless 
parameter $\xi $ implies that the response of the system is in general a nonlinear function 
of the magnitude of the static component $E^{(0)}$. This dependence also influences the 
characteristic relaxation time as known from experiments \cite{block}.


Our formalism reduces to well known phenomenological relations used to
describe dielectric relaxation in diverse materials including solids,
glass-forming liquids and polymer melts~\cite{jonscher001}, when the 
appropriate limiting cases are considered. 

In particular, Eqs.~(\ref{illustration010}) and (\ref{illustration011}) can be
reduced to the Debye and Cole-Cole~\cite{cole001} dielectric response functions if 
the infinite hierachy of dynamic equations for the
moments of the distribution, starting with Eqs.~(\ref{perturbation003}) and 
(\ref{perturbation004}), is truncated under the assumption that the order
parameter tensor, $S_{ij}$, remains very close to its equilibrium value, i.e.
$S_{ij}^{(0)}=S_{ij}^{\text{eq}}\left( \vec{E}^{(0)}\right)$, and
$S_{ij}^{(1)}=0$.  If in addition we neglect the diffusion of polarization by assuming
$\varphi \left( t \right)= 0$, for all $t$, then Eqs.~(\ref{illustration010}) and
(\ref{illustration011}), reduce to the simpler form
\begin{equation}
\chi _{\perp ,\parallel }\left( \omega \right) =\chi _{\perp
,\parallel }^{o}\frac{2\phi \left( \omega \right) }{i\omega +2\phi \left(
\omega \right) },  \label{limiting001}
\end{equation}%
where the amplitudes of the parallel and perpendicular components, $\chi
_{\parallel }^{o}$ and $\chi _{\perp }^{o}$, read \cite{mendez001}
\begin{equation}
\chi _{\parallel }^{o}=3\chi _{0}\frac{L\left( \xi \right) }{\xi }\text{,
and }\chi _{\perp }^{o}=\frac{3}{2}\chi _{0}\left( 1-\frac{L\left( \xi
\right) }{\xi }\right).  \label{limiting002}
\end{equation}%
When the response of the system can be considered instantaneous, we have
$\phi \left( t \right) =D \delta \left( t\right)$, i.e. 
$\phi \left( \omega \right) =D$, and therefore Eq.~(\ref{limiting001}) takes the form of 
two independent Debye expressions
\begin{equation}
\chi _{\perp ,\parallel }\left( \omega \right) =\chi _{\perp
,\parallel }^{o}\frac{1}{1+i\omega \tau_{0} },  \label{limiting003}
\end{equation}%
where the relaxation time, $\tau_{0}$, is related with the rotational diffusion
coefficient through $\tau_{0} =1/2D$. 
On the other hand, for the power law memory kernel given by Eq.~(\ref{frac002}),
we have $\phi \left( \omega\right) =D^{\alpha }\left( i\omega \right) ^{1-\alpha }$, 
and Eq.~(\ref{limiting001}) reduces to two expressions of the Cole-Cole type
\begin{equation}
\chi _{\perp ,\parallel }\left( \omega \right) =\chi _{\perp
,\parallel }^{o}\frac{1}{1+\left( i\omega \tau _{cc}\right) ^{\alpha }},
\label{limiting004}
\end{equation}%
where the relaxation time appearing in this expression, $\tau _{cc}$, is
given by $\tau _{cc}=1/2^{1/\alpha }D$.

In the following subsections we will analyze in more detail the 
dependence on the frequency, the normalized dipole electrostatic energy, the 
rotational diffusion and the diffusion of polarization of the real and imaginary parts of the
complex susceptibility, $\chi _{\perp ,\,\parallel }^{\prime }=\text{Re}%
\left\{ \chi _{\perp ,\,\parallel }\right\} $ and $\chi _{\perp ,\,\parallel
}^{\prime \prime }=-\text{Im}\left\{ \chi _{\perp ,\,\parallel }\right\} $.

\subsection{The case of Markovian dynamics}

For simplicity, in the following the analysis will be carried out in different 
successive reduced levels of description in order to better quantify and elucidate 
the effects due to couplings, diffusion of polarization and the non-Markovian character 
of the system. With this aim we have subdivided this section into subsections.

\subsubsection{Effects due to the coupling with higher order moments}

\begin{figure}[tbp]
\includegraphics[width=80mm]{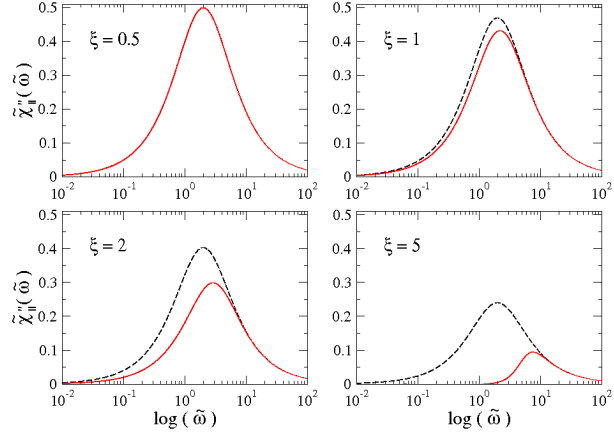}
\caption{Imaginary part of the parallel component of the dielectric
susceptibility for diverse values of the external field normalized with 
respect to the average thermal energy, $\xi$. All curves correspond
to Markovian processes in the homogeneous approximation. Solid lines
contain the effect of the coupling of high order moments of the distribution
function as obtained from Eq.~(\protect\ref{illustration011}) with 
$\tilde{G}_{1}$ and $\tilde{G}_{2}$ given by Eqs.~(\protect\ref{markovhomo001}) and 
(\protect\ref{markovhomo002}). Dashed lines are obtained from the imaginary 
part of Eqs.~(\protect\ref{limiting001}) and (\ref{limiting002}), in which the 
coupling with high order moments is not taken into account (Debye case).}
\label{figure001}
\end{figure}

\begin{figure}[tbp]
\includegraphics[scale=0.50]{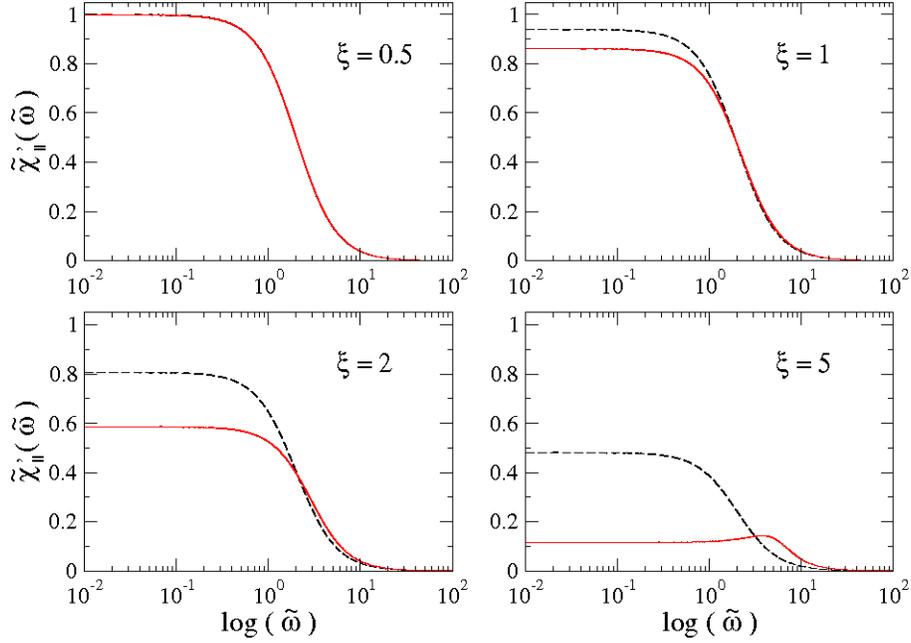}
\caption{The same as in Fig.~\ref{figure001} for the real part of the 
parallel component of the dielectric susceptibility. Solid lines represent the 
effect of the coupling of high order moments whereas the dashed lines correspond 
to the Debye case.}
\label{figure002}
\end{figure}

We will consider first the effects of the coupling of the average polarization
with the moments of second and third order on the dielectric susceptibility. 
From Eqs.~(\ref{perturbation008a}) and (\ref{perturbation008b}), 
and~(\ref{illustration010})-(\ref{limiting002}), 
it follows that coupling effects are manifested by the terms proportional to 
$\tilde{G}^{-1}_{2}\left( \vec{k},\omega \right)$, which is the propagator associated
with the time and space evolution of $S_{ij}$. Ignoring for convenience 
the effects associated to diffusion of polarization by
making $\varphi \left( \omega \right) = 0$, and the non-Markovian effects
by assuming an instantaneous rotational diffusion kernel, i.e. 
$\phi \left( \omega \right) = D$, the normalized 
complex susceptibilities perpendicular and parallel to $\vec{E}^{(0)}$, 
Eqs.~(\ref{illustration010}) and (\ref{illustration011}), are given by 
\begin{equation}
\tilde{G}_{1}\left( \vec{k},\tilde{\omega}\right) = 1 + i\tilde{\omega},
\label{markovhomo001}
\end{equation}
\begin{equation}
\tilde{G}_{2}\left( \vec{k},\tilde{\omega}\right) = 3 + i\tilde{\omega},
\label{markovhomo002}
\end{equation}
where the normalized frequency, $\tilde{\omega}$, has been defined  by
\begin{equation}
\tilde{\omega}=\omega /2D.  \label{markovhomo003}
\end{equation}
When these expressions for $\tilde{G}_{1}$ and $\tilde{G}_{2}$ are used in 
Eqs.~(\ref{illustration010}) and (\ref{illustration011}), the result will contain 
exclusively the effects of the dynamics of the second and third order moments on 
$\chi_{\perp}$ and $\chi_{\parallel}$. This result has to be compared with 
Eq.~(\ref{limiting003}), in which these effects have been neglected. In 
Figs.~\ref{figure001} and \ref{figure002} we present this comparison for the real
and imaginary parts of the normalized susceptibility  
$\tilde{\chi}_{\parallel }\left( \vec{k},\omega \right) 
= \chi _{\parallel }\left( \vec{k},\omega \right) /\chi _{0}$, and four values of 
the normalized electrostatic energy, $\xi $, ranging from small, $\xi =0.5$, to 
large values, $\xi =5$, as compared with the average thermal energy per molecule. 
It can be observed that 
when the external static field is small, the coupling with high order moments produce no 
significant effect on the the complex susceptibility. However, for moderate, $\xi = 1.0$, 
and large external static fields, $\xi = 5$, the polar moments of the distribution increase 
and the changes induced by the coupling between these moments in both 
$\chi^{\prime} \left( \omega \right)$ and $\chi^{\prime \prime} \left( \omega \right)$, 
are appreciable. 

Figs.~\ref{figure001} and \ref{figure002} show that the coupling of higher order moments 
reduces the susceptibility of the system with respect to the Debye-like case, in which this
coupling is neglected. It can be also noticed that the frequency at which 
$\chi^{\prime \prime}_{\parallel}$ takes its maximum value is shifted towards the region of
large frequencies. However, the frequency dependence of 
$\chi^{\prime \prime}_{\parallel} \left( \omega \right)$ is not modified in this limit since
this quantity still decreases as $\omega^{-1}$. 



\subsubsection{Effects due to diffusion of polarization}

The diffusion of polarization may also influence dependence on the wave vector and 
frequency of the dielectric susceptibility as we will show in this section. In order to 
focus on these effects, in this case we will consider the Markovian dynamics with memory 
functions $\phi \left( \omega \right) = D$, for the rotational diffusion, and  
$\varphi \left( \omega \right) =\mathcal{D}$, for the diffusion of polarization, 
that has associated the diffusion  coefficient $\mathcal{D}$. We will also ignore the 
effects due to the coupling of the average dipole moment with moments of second and third 
order by making $\tilde{G}^{-1}_{2}=0$ in Eqs.~(\ref{illustration010}) 
and (\ref{illustration011}). 

Within these approximations, only the normalized propagator $\tilde{G}_{1}$ is 
required for describing the behavior of $\chi_{\perp,\parallel}$, and it
can be written in the form
\begin{equation}
\tilde{G}_{1}\left( \vec{k},\tilde{\omega}\right) =1+\gamma +i\tilde{\omega},
\label{markovnonhomo001}
\end{equation}
where the parameter $\gamma =k^{2}\mathcal{D}/2D$, has been introduced to
quantify the relative contribution of the diffusion of polarization with
respect to rotational diffusion of the molecules. The normalized frequency $\tilde{\omega}$
is given as before by Eq.~(\ref{markovhomo003}).

When Eq.~(\ref{markovnonhomo001}) together with $\tilde{G}^{-1}_{2} = 0$, are
used in Eqs.~(\ref{illustration010}) and (\ref{illustration011}), we obtain
expressions which exclusively quantify the effects of the diffusion of polarization
on the components of the dielectric susceptibility. Since in this approximation the
diffusion of polarization only modifies the real part of $\tilde{G}_{1}$, it 
is clear that its general effect on the imaginary part of $\chi_{\perp}$ and 
$\chi_{\parallel}$ will consist in decreasing the value and shifting the location 
of the maxima of these functions in an amount proportional to the parameter 
$\gamma$. However, it can be noticed that the general dependence of the dielectric 
susceptibility with respect to the frequency is not modified by a memory kernel
of the form $\varphi = \mathcal{D}$, because both $\chi^{\prime \prime}_{\perp}$
and $\chi^{\prime \prime}_{\parallel}$ increase as $\omega$, 
for $\tilde{\omega} \ll 1 + \gamma$, and decrease as $\omega^{-1}$, for
$\tilde{\omega} \gg 1 + \gamma$. Similar conclusions are valid for the real 
part of $\chi_{\perp}$ and $\chi_{\parallel}$.

\subsection{The case of non-Markovian dynamics}

In this section, we will first consider the effects of the non-instantaneous
response of the system to the imposed fields and afterwards both, the effects 
coming from coupling with higher order moments and those due to the diffusion 
of polarization. Consequently, we set  $\varphi \left( \omega \right) = 0$, 
and $G^{-1}_{2} = 0$, in Eqs.~(\ref{illustration010}) and (\ref{illustration011}). 
This yields Eqs.~(\ref{limiting001}) and (\ref{limiting002}) for the dielectric 
susceptibility components, in which the memory kernel, $\phi \left( \omega \right)$, 
has to be given in order to specify the behavior of the dielectric relaxation.

For the sake of clarity, in the present paper, we will only analyze the effects 
of two types of memory kernels, namely, the power law memory function given by 
Eq.~(\ref{frac002}), and a modified kernel that combines both exponential and 
power law decay. We will write the latter in the form 
\begin{equation}
\phi \left( t\right) =
\left\{ 
\begin{array}{cc}
\frac{2 D^{2}}{ \Gamma \left( \alpha -1\right) }
e^{-2Dt} \left( 2 D t \right) ^{\alpha -2} & 
\text{for }t>0 \\ 
0 & \text{for }t<0
\end{array}
\right. ,  \label{homo000}
\end{equation}
with $0<\alpha <1$.

Let us discuss first the effects of the memory function given by 
Eq.~(\ref{frac002}). Previously, we have shown that this function 
yields the FFPE, Eq.~(\ref{frac006}), and the Cole-Cole law for the 
dielectric susceptibility, Eq.~(\ref{limiting004}). Accordingly, 
the effects of this kind of non-Markovian dynamics can be analyzed by simply 
comparing our previous results, namely, Eq.~(\ref{limiting004}) with 
respect to the Debye expression, Eq.~(\ref{limiting002}), in which 
such effects do not appear. 

\begin{figure}[tbp]
\includegraphics[scale=0.50]{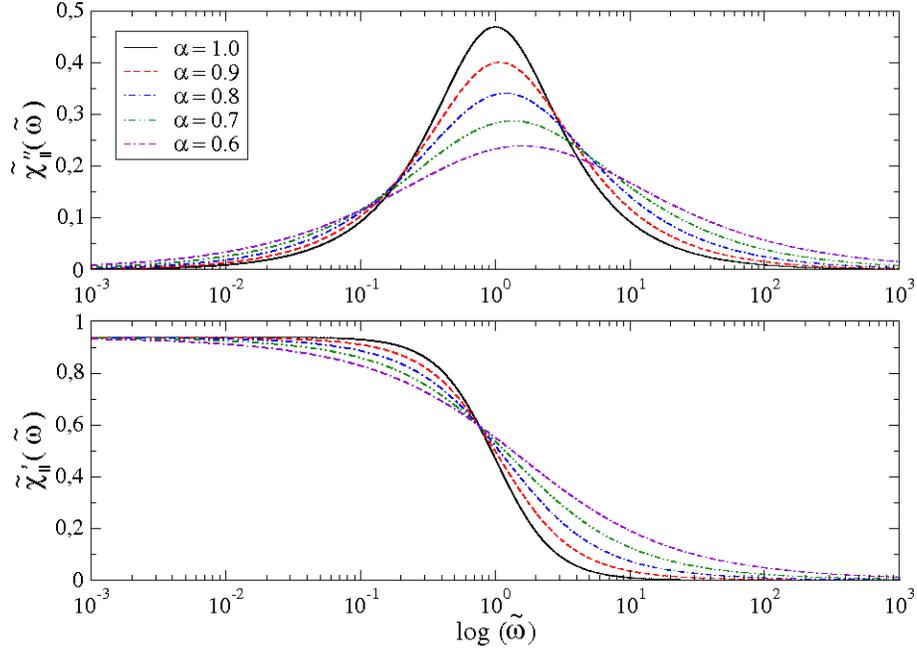}
\caption{Normalized parallel component of the dielectric susceptibility
in the non-Markovian homogeneous approximation. The memory kernel 
describing the rotational diffusive processes is considered to decay
according to the power law Eq.~(\protect\ref{frac002}). The dimensionless
exponent $\protect\alpha$, takes diverse values including $\protect\alpha=1$,
which corresponds to normal diffusion, and $\protect\alpha=0.9,0.8,0.7,0.6$, 
which describe rotational subdiffusive processes. 
$\tilde{\chi}_{\parallel }^{\prime \prime}$ and $\tilde{\chi}_{\parallel }^{\prime}$
are calculated from Eqs.~(\protect\ref{limiting004}) and (\protect\ref{limiting003}), 
with $\tilde{\chi}_{\parallel}=\chi_{\parallel}/\chi_{0}$. The value
of the normalized field is taken as $\xi = 1$.}
\label{figure003}
\end{figure}

In Fig.~\ref{figure003} we perform this comparison for the
real and imaginary parts of $\chi_{\parallel} \left( \omega \right)$.
For convenience, we have illustrated our results in terms of the 
normalized susceptibility 
$\tilde{\chi}_{\parallel}=\chi_{\parallel}/\chi_{0}$, and the normalized
frequency $\tilde{\omega}$, Eq.~(\ref{markovhomo003}). 

Curves in Fig.~\ref{figure003} exhibit the well known features of the
Cole-Cole susceptibility. The dielectric peak in the Cole-Cole case is
symmetric and behaves in the form
$\chi^{\prime \prime} \sim \omega^{\alpha}$,
$\chi^{\prime \prime} \sim \omega^{-\alpha}$, in the limits of small and
large frequencies, respectively.
For $ 0 < \alpha <1$, the width of the dielectric loss increases 
the smaller $\alpha$ is, while the departure with respect to the 
Debye case is eliminated for $\alpha = 1$. 
The frequency dependence of the real part of the susceptibility 
is also modified by  the power law memory function, Eq.~(\ref{frac002}),
and instead of the characteristic Debye's decay, 
$\chi^{\prime} \sim \omega^{-2}$, a decay of 
the form $\chi^{\prime} \sim \omega^{-\alpha} $,  is observed.


The modified memory function given by Eq.~(\ref{homo000}) yields
an asymmetric shape for the dielectric peak. This result follows from the 
Fourier transform of Eq.~(\ref{homo000}): 
$\phi \left( \omega \right) = D \left( 1 + i \omega / 2D \right)^{1-\alpha}$,
and after replacing this expression into Eq.~(\ref{limiting001}). The 
resulting formula for the normal and parallel components of the susceptibility 
can be written in the form 
\begin{equation}
\chi_{\perp , \parallel} \left( \omega \right)
= \frac{\chi_{\perp , \parallel}^{o} }
{1 + i \omega \tau_{0} \left( 1 + i \omega \tau_{0} \right)^{\alpha-1}}
, \label{homo003}
\end{equation}
where we used $\tau_0=1/2D$ and the amplitudes $\chi_{\perp , \parallel}^{o}$ 
are given by Eq.~(\ref{limiting002}). 

\begin{figure}[tbp]
\includegraphics[scale=0.50]{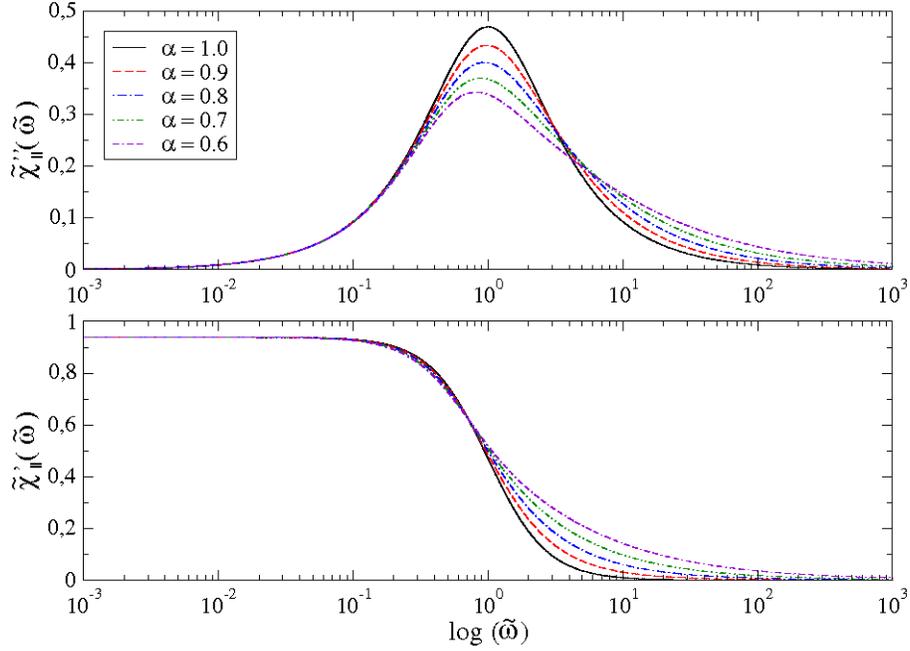}
\caption{Normalized parallel component of the dielectric susceptibility
in the non-Markovian homogeneous approximation. The memory kernel
describing the rotational diffusive processes is considered to decay
according to the modified power law Eq.~(\protect\ref{homo000}). For
$\protect\alpha=1$, the usual decay in the dielectric loss 
$\tilde{\chi}^{\prime \prime}_{\parallel} \sim \tilde{\omega}^{\pm 1}$, 
is observed. For $\alpha < 1 $, $\tilde{\chi}^{\prime \prime}$ becomes
asymmetric, increasing as $\tilde{\omega}$ and decreasing as $\tilde{\omega}^{-\alpha}$.
$\tilde{\chi}_{\parallel }^{\prime \prime}$ and $\tilde{\chi}_{\parallel }^{\prime}$
are calculated from Eqs.~(\protect\ref{homo003}) and (\protect\ref{limiting003}),
with $\tilde{\chi}_{\parallel}=\chi_{\parallel}/\chi_{0}$. The value
of the normalized field is fixed at $\xi = 1$.}
\label{figure004}
\end{figure}

In order to analyze the low and high frequency behavior of Eq. (\ref{homo003}), 
we will use the formula
\begin{equation}
\left( 1 + i \omega \tau_{0} \right)^{\alpha-1} =
\left( 1 + \omega^{2} \tau^{2}_{0} \right)^{\frac{\alpha-1}{2}}
e^{i\left( \alpha-1 \right) \tan^{-1}\left( \omega \tau_{0} \right) }. \nonumber
\end{equation}
In the limit $\omega \tau_{0} \ll 1$, the imaginary part of 
$\chi_{\perp , \parallel}$ increases linearly with $\omega$,
i.e.  $\chi_{\perp , \parallel} \sim \omega$; while for $\omega \tau_{0} \gg 1$,
it decreases as $\chi_{\perp , \parallel} \sim \omega^{-\alpha}$. This asymmetry
in the dielectric loss is not exhibited by the Debye nor by the Cole-Cole 
functions, but resembles the dependence that is observed in systems described
by the phenomenological Cole-Davidson expression~\cite{jonscher001}.
In a similar way, it can be shown that the real part of
$\chi_{\perp , \parallel}$ decreases as 
$\chi_{\perp , \parallel} \sim \omega^{-\alpha}$ for $\omega \tau_{0} \gg 1$, 
in contrast with the usual Debye dependence. Fig.~\ref{figure004} exhibits 
these effects on the real and imaginary parts of 
$\chi_{\parallel} \left( \omega \right)$, in terms of the normalized susceptibility
$\tilde{\chi}_{\parallel}=\chi_{\parallel}/\chi_{0}$, and the normalized
frequency $\tilde{\omega}$, Eq.~(\ref{markovhomo003}).

The general case in which both, rotational diffusion and the diffusion of 
polarization processes are present and have a non-Markovian character can 
be analyzed as a function of the parameter $\xi$. In general terms, the
qualitative correction arising from the coupling with higher orders moments 
only modulates the amplitudes of the obtained functions and their 
characteristic relaxation times (their location along the $\omega$-axis), 
as shown in Figs.~\ref{figure001} and \ref{figure002}. Thus, if for 
simplicity we consider small values of $\xi$, such that the coupling of high
order moments is neglected, then the expression for the components of 
$\chi \left( \vec{k}, \omega \right)$ takes the form
\begin{equation}
\chi _{\perp ,\parallel }\left( \omega \right) 
=\chi _{\perp,\parallel }^{o}
\frac{2\phi \left( \omega \right) }
{i\omega +2\phi \left( \omega \right) + k^{2} \varphi\left( \omega \right) },  
\label{homo005}
\end{equation}
with $\chi_{\perp,\parallel }^{o}$ given by Eq.~(\ref{limiting002}). When
the memory kernels $\phi \left( t \right)$ and $\varphi \left( t \right)$
are assumed to decay as power laws in time, Eqs.~(\ref{frac002}) and 
(\ref{frac007}), then Eq.~(\ref{homo005}) reduces to
\begin{equation}
\chi _{\perp ,\parallel }\left( \omega \right) 
=
\frac{ \chi _{\perp,\parallel }^{o} }
{ 1 + \frac{1}{2} \left( \frac{i\omega}{D} \right)^{\alpha} 
+ \frac{ k^{2} \mathcal{D}_{\beta} D^{-\beta}}{2} 
\left( \frac{i\omega}{D}\right)^{\alpha -\beta} } ,  
\label{homo006}
\end{equation}
which, as Eq.~(\ref{homo003}), is also asymmetric. Unlike in the previous 
case, now the low and high frequency behaviors of the real and imaginary 
parts of the susceptibilities will be determined by the exponents $\alpha$ 
and $\beta$. This result implies that the diffusion of polarization 
introduces asymmetries in the response functions of the system that depend 
on the viscoelastic properties of the medium. The representation of 
Eq.~(\ref{homo006}) contains a combination of the effects shown in 
Figs.~\ref{figure003} and \ref{figure004}.

\subsection{Cooperative effects}

A simple and effective generalization of the present model can be formulated 
along the following lines. Consider the case of a system in which different 
kinds of polar elements, characterized by their dipolar moments, 
$p^{(q)}$, are present. In this case, the probability distribution function will 
depend on several orientation vectors, $\vec{n}^{(q)}$, associated to each 
type of polar element. For example, in the case $q=2$, the distribution 
function $f(\vec{r}^{(1)},\vec{r}^{(2)},\vec{n}^{(1)},\vec{n}^{(2)}, t)$, will 
satisfy a bivariate conservation equation entirely similar to Eq.~(\ref{mnet001}), 
but having the corresponding divergence term in the 
$(\vec{r}^{(1)},\vec{r}^{(2)},\vec{n}^{(1)},\vec{n}^{(2)})$-space. 

As a consequence, a bivariate Fokker-Planck equation similar to Eq.~(\ref{mnet009}) 
can be obtained from the mesoscopic thermodynamics analysis. This equation
can in turn be used to formulate  the corresponding evolution equations for 
the moments of the distribution, in which 
$P^{(q)}_{i}(\vec{r},t)=%
\mathcal{N}_{q} \int d\vec{n}^{(1)} d\vec{n}^{(2)} d\vec{r}^{(1)} d\vec{r}^{(2)} 
p^{(q)}_{i} 
f \delta(\vec{r}-\vec{r}^{(q)}) $, $q=1,2$,
are the dipole moments, see Eq.~(\ref{multi001}). Here,  ${\cal N}_q$ is the 
number of polar elements of class $q$, and the operator $\delta(\vec{r}-\vec{r}^{(q)})$, 
evaluates the value of the field $\vec{P}^{(q)}$ in the continuous position $\vec{r}$, 
see for example~\cite{ivanPRE}. The definitions of the quadrupolar and octupolar 
moments are therefore given by entirely similar relations as those of 
Eqs.~(\ref{multi002}) and (\ref{multi003}). 

If we assume that these polar elements are independent, a simple and 
powerful model can be formulated to account for experiments. Under the 
independence assumption, the evolution equations associated with 
$\vec{P}^{(1)}(\vec{r},t)$ and $\vec{P}^{(2)}(\vec{r},t)$, are also independent 
and therefore, the total relaxation process results from the sum of both 
contributions. Taking into account only  
the time dependent field, the relation between the 
polarization and the applied electric field can be written as
\begin{equation}
P_{i} = \sum_{m=1}^{q} P^{(m)}_{i}( \vec{k},\omega) 
             =\chi_{ij}(\vec{k},\omega)  E_{j}(\vec{k},\omega) ,  
\label{Many001}
\end{equation}
where $\chi_{ij}(\vec{k},\omega) =\sum_{m=1}^{q}\chi^{(m)}_{ij}( \vec{k},\omega)$,
since in the present approximation the response function is independent of 
the applied field.

As an example, consider that $q=2$ and that the diffusion of polarization 
effect is subdiffusive anomalous. Thus, by applying a similar procedure to 
the one yielding Eq.~(\ref{homo006}), we obtain
\begin{equation}
\chi\left( \omega \right) 
= \frac{ \chi _{1}^{o} }
{ 1 + \frac{1}{2} \left( \frac{i\omega}{D_{1}} \right)^{\alpha_{1}} 
+ \frac{ k^{2} \mathcal{D}_{\beta_{1}} D_{1}^{-\beta_{1}}}{2} 
\left( \frac{i\omega}{D_{1}}\right)^{\alpha_{1} -\beta_{1}} } + \frac{ \chi _{2}^{o} }
{ 1 + \frac{1}{2} \left( \frac{i\omega}{D_{2}} \right)^{\alpha_{2}} 
+ \frac{ k^{2} \mathcal{D}_{\beta_{2}} D_{2}^{-\beta_{2}}}{2} 
\left( \frac{i\omega}{D_{2}}\right)^{\alpha_{2} -\beta_{2}} } ,  
\label{Many002}
\end{equation}
where the subscripts for the parallel and perpendicular components have
been suppressed in order to simplify the notation.
In this case, the susceptibility is also asymmetric but it may present 
two broad peaks when the characteristic relaxation times associated to 
$\vec{P}^{(1)}$ and $\vec{P}^{(2)}$ are sufficiently separated. 

\begin{figure}[tbp]
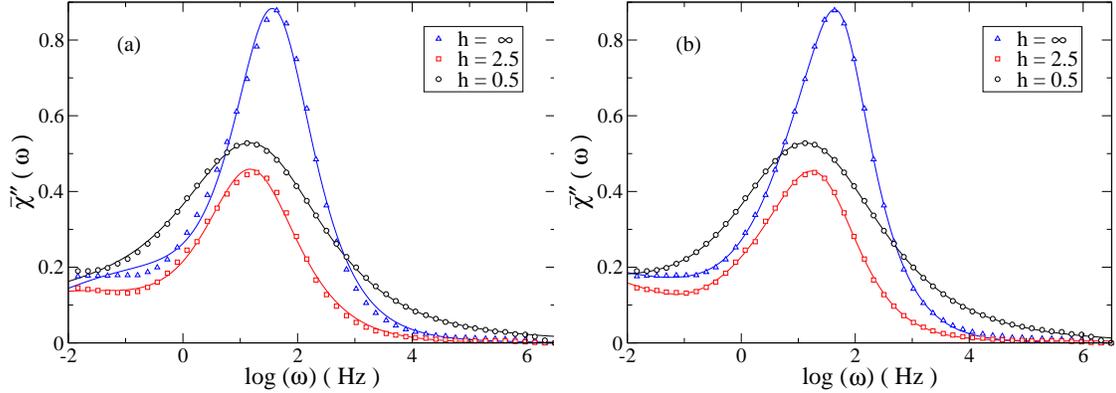

\includegraphics[scale=0.30]{figure005a.eps}
\includegraphics[scale=0.30]{figure005b.eps}
\caption{Normalized dielectric loss of glycerol/water samples at 160$K$ in presence 
of myoglobin with different levels of hydration. The symbols represent 
experimental results taken from Ref.~\cite{PNASMb} whereas the lines are 
the fits from Eq.~(\ref{Many002}). Our results suggest that the structural 
$\alpha$-relaxation process affects the observed $\beta$-relaxation
spectrum by introducing an asymmetry in the amplitudes of the spectrum at 
low frequencies. }
\label{figure005}
\end{figure}

The imaginary part of Eq.~(\ref{Many002}) can be used to account for the 
dielectric relaxation spectra of glassy materials. This is shown in 
Fig.~\ref{figure005}, where a comparison between experiments (symbols) 
and theory (lines) is
presented in a normalized representation of the dielectric loss,
 $\bar{\chi}^{\prime \prime}$, see Ref.~\cite{PNASMb}. The experiments were performed with 50/50 (wt/wt) 
glycerol/water samples at 160~K (solid), in the absence and the presence 
of myoglobin at different levels of hydration. 
The symbol $h$ stands for the level of hydration, where $h = \infty$ corresponds 
to the case when no myoglobin was added. The dielectric relaxation spectroscopy 
experiments were performed by introducing the samples in a capacitor and applying 
an oscillatory electric field~\cite{PNASMb}. The presence of the hydrated myoglobin 
broadens the spectrum and shifts the maximum towards low frequencies. 
In Fig.~\ref{figure005} we have included two cases: (a) Without considering
diffusion of polarization, i.e. $\mathcal{D}_{\beta_{1}}=\mathcal{D}_{\beta_{2}} = 0$, 
in Eq.~(\ref{Many002}). (b) Considering the diffusion of polarization, i.e.  
$\mathcal{D}_{\beta_{1}} \neq \mathcal{D}_{\beta_{2}} \neq 0$.  It can be observed that 
the diffusion of polarization improves the agreement between theoretical and experimental results. 

It is interesting to notice that the diffusion of polarization appears 
to have most dramatic effects in the absence of myoglobin, while its effects 
are small in the presence of myoglobin. These effects can be understood by 
considering that the reduction of the amplitude of the dielectric loss is due 
to the presence of both the myoglobin and the hydration shell, whereas the 
addition of new modes at low and high frequencies can be attributed to the 
presence of the hydration shell~\cite{PNASMb}. Then, it is plausible to assume 
that the hydration shell induces long-ranged orientational correlations between 
solvent particles, which reduce the relative importance of the 
diffusion of polarization term with respect to the non-diffusive terms in all 
the evolution equations for the multipolar moments. The orientational correlations 
explain why in Fig.~\ref{figure005}a the agreement for the larger value of the 
hydration number ($h=2.5$) is better at low frequencies than in the case of low 
hydration numbers ($h=0.5$). For low $h$, the hydration shell is less important 
and therefore the amplitude of the dielectric loss increases as well as the 
importance of the diffusion term in the evolution equations for the multipoles.

In this specific application, the superposition of two independent
dielectric responses given by Eq.~(\ref{Many002}), accounts very well for
the complicated shape of the dielectric loss found experimentally.
According to our model, this function should consist of a peak centered at 
small frequencies, associated with the so called $\alpha$-relaxation processes, 
superposed to a peak located at larger frequencies, which is associated
with slower or $\beta$-relaxation processes. The latter peak is the one 
observed in Figs.~\ref{figure005}~(a) and (b). It can be noticed that
the diffusion of polarization allow for fitting the experimental data
in the range of small frequencies. Notice that, as $h$ increases, 
the exponent $\beta$ increases as well, implying that $\beta (h)$ and 
indicating that hydration favors a transition from anomalous to normal 
diffusion of polarization, see the tables in Appendix~B.

\section{Discussion}

A physical insight of the mechanism underlying the power-law behavior of 
the memory kernel in Eq.~(\ref{frac002}) can be elucidated by decomposing 
the corresponding spectrum in terms of elementary Debye processes, that is, 
by appealing to the theory of distribution of relaxation times~\cite{balescu}.
For simplicity, our analysis will be focused on the description leading to Cole-Cole 
parallel susceptibility (\ref{limiting004}) with the use of the power law 
memory function, Eq.~(\ref{frac002}). 

We will proceed by assuming the existence of a distribution of elementary
processes with proper relaxation times and amplitudes, whose superposition 
generates a wider spectrum such that corresponding to the Cole-Cole function, 
Eq.~(\ref{limiting004}). In this form, the Cole-Cole function can be recovered 
from a superposition of elementary Debye processes through the 
relation~\cite{bello001}
\begin{equation}
\chi_{\parallel} \left( \omega \right) 
 = \int_{-\infty}^{\infty}d\left( \ln |\tilde{\tau}|\right)
  \psi\left(\ln |\tilde{\tau} | \right) 
 \frac{1}{1+i\omega \tau} 
 = \int_{-\infty}^{\infty}dy\,
  \psi\left( y \right) 
 \frac{1}{1+i\omega \tau_{cc}e^{y}} ,
\label{limiting008}
\end{equation}
where for simplicity in notation we have introduced the variable 
$\tilde{\tau}= {\tau}/{\tau_{cc}}$ and used the definition 
$y=\ln| \tilde{\tau} |$ in the second equality. The function 
$\psi\left(y \right) $ is a continuous distribution of relaxation 
times having the explicit form
\begin{equation}
\psi\left( y\right) 
= \frac{\chi^{o}_{\parallel}}{2 \pi}
  \frac{  \sin  \left( \alpha \pi\right)}
       {\cosh \left( \alpha y \right)
        +\cos\left( \alpha\pi \right)} 
. \label{limiting009}
\end{equation}

In fact, the analysis can be simplified by considering  a discrete
expansion of Eqs.~(\ref{limiting008}) and (\ref{limiting009}), that is,
by approximating the Cole-Cole response law by a superposition of
independent Debye-like processes $\chi_{k}\left( \omega \right)$, distributed
over a discrete space in the form
\begin{equation}
\chi_{\parallel} \left( \omega \right) =
\sum_{k=-\infty}^{\infty} \chi_{k} \left( \omega \right) =
\sum_{k=-\infty}^{\infty} \frac{A_{k}}{1+i\omega \tau_{k}},
\label{limiting010}
\end{equation}
where the amplitudes, $A_{k}$, and relaxation times, $\tau_{k}$, of the
$k$-th process are given, respectively, by
\begin{equation}
A_{k} = \frac{\chi^{o}_{\parallel}}{2 \pi}
  \frac{  \sin  \left( \alpha \pi\right)}
       {\cosh \left( \alpha k \right)
        +\cos\left( \alpha\pi \right)},
\text{ and } 
\tau_{k} = \tau_{cc} e^{k} .
\label{limiting011}
\end{equation}
Notice that the index $k$ in the summation of Eq.~(\ref{limiting011}),
quantifies the time-scale separation between the $k$-th elementary process
and the process with the relaxation time $\tau_{cc}$, which defines the 
location of the maximum of the dielectric response function. Those processes 
labeled by an index $k \gg 1$, evolve  toward equilibrium in
a characteristic time which is much larger than $\tau_{cc}$, while those with
$k \ll -1$, decay much faster than the processes with relaxation time of
the order of $\tau_{cc}$.

\begin{figure}[tbp]
\includegraphics[scale=0.50]{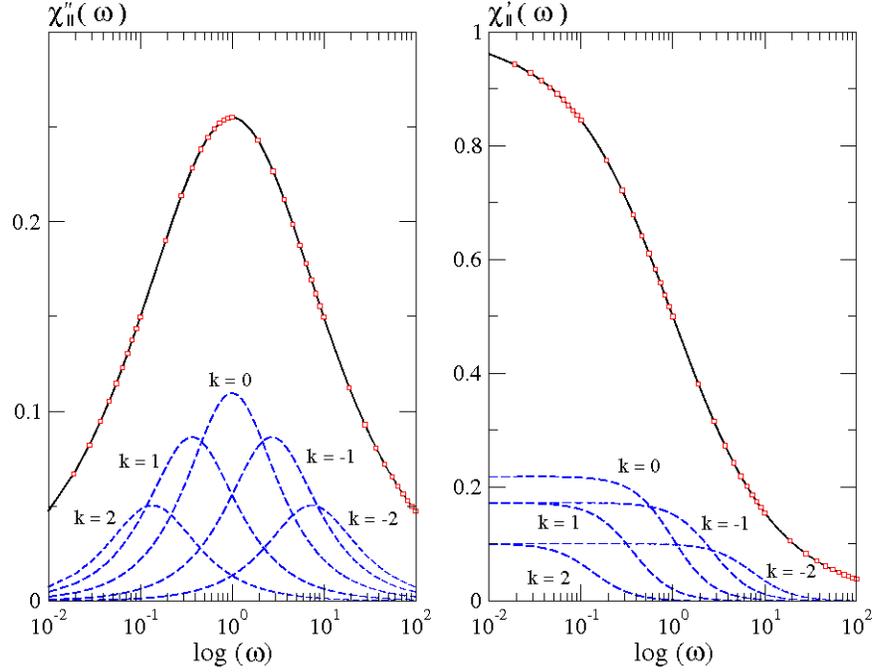}
\caption{Approximation of a Cole-Cole function by a superposition of
discrete Debye components with amplitudes and relaxation times given by
Eq.~(\ref{limiting011}). The real and imaginary part of the Cole-Cole 
function, Eq.~(\ref{limiting004}), are represented by the continuous curve. 
They correspond to $\alpha = 0.6$. The superposition of Debey components,
Eqs.~(\ref{limiting010}) and (\ref{limiting011}),
is represented by the square symbols and it can be observed that
it fits the Cole-Cole function over the whole range of frequencies.
Some of the Debye components, those with indexes $k=-2,-1,0,1,2$, are 
explicitly shown.}
\label{figure006}
\end{figure}

Fig.~\ref{figure006} illustrates the approximation of both the real and
imaginary parts of the normalized Cole-Cole function, 
\begin{equation}
\frac{\chi^{\prime}_{\parallel} \left( \omega \right) }{\chi^{o}_{\parallel}}
=
  \frac{1+\left( \tau_{cc} \omega \right)^{\alpha} \cos \left( \alpha \pi / 2\right)}
       {\left[ 1+\left( \tau_{cc} \omega \right)^{\alpha} \right]^{2}
        + \left( \tau_{cc} \omega \right)^{2\alpha} \sin^{2} \left( \alpha \pi / 2\right)}
, \label{limiting011a}
\end{equation}
\begin{equation}
\frac{\chi^{\prime \prime}_{\parallel} \left( \omega \right)}{\chi^{o}_{\parallel}}
= 
  \frac{ \left( \tau_{cc} \omega \right)^{\alpha} \sin \left( \alpha \pi / 2\right)}
       {\left[ 1+\left( \tau_{cc} \omega \right)^{\alpha} \right]^{2}
        + \left( \tau_{cc} \omega \right)^{2\alpha} \sin^{2} \left( \alpha \pi / 2\right) }
, \label{limiting011b}
\end{equation}
by the superposition of Debye components over the discrete
distribution of relaxation times given by Eq.~(\ref{limiting011}) with $\alpha=0.6$,
and where the infinite sum in Eq.~(\ref{limiting011}) has been approximated
by a sum over a finite interval ranging from $k=-100$ to $k=100$.
Eq.~(\ref{limiting011}) constitutes a very good approximation to the Cole-Cole function, reproducing
very well the behavior of the latter over the whole frequency range considered.  
It is interesting to notice that a partial superposition of independent Debye contributions in which 
elementary processes within a given range are suppressed, yields a two step relaxation dynamics
manifested through the dielectric loss with well marked peaks as those observed 
experimentally~\cite{kotaka}. This elimination of elementary processes 
can be represented by non-uniform distributions of characteristic times~\cite{hijar001}.

These results can be used to give an insight of the physical mechanisms 
underlying the relaxation process if we consider
Eqs.~(\ref{limiting002}) and (\ref{limiting003}),
and the definition of the static susceptibility 
$\chi_{0} = \mathcal{N}p^{2}/3k_{B}T$. The amplitude of each elementary Debye 
contribution in Eq.~(\ref{limiting010}), can be written in terms of an effective
numerical density of polar elements $\mathcal{N}_{k}$ participating in the process 
with relaxation time $\tau_{k}$, that is
\begin{equation}
A_{k} = \frac{\mathcal{N}_{k} p^{2}}{k_{B}T}\frac{L\left( \xi \right)}{\xi}.
\label{limiting013}
\end{equation}

Comparing Eqs.~(\ref{limiting011}) and (\ref{limiting013}) we obtain
\begin{equation}
n_{k} = \frac{\xi}{6\pi L\left( \xi \right)} 
        \frac{ \sin  \left( \alpha \pi\right)}
             {\cosh \left( \alpha k \right)
              +\cos\left( \alpha\pi \right)}
, \label{limiting014}
\end{equation}
where $n_{k}=\mathcal{N}_{k}/\mathcal{N}$, is the effective fraction of polar elements
which contribute to process $k$. In the limit of slow and fast processes, i.e. $k \gg 1$ 
and $k \ll -1$, respectively, and in terms of the relaxation time $\tau_{k}$, 
Eq.~(\ref{limiting014}), $n_{k}$ can be recast into
\begin{equation}
n_{k} \left( \tau_{k} \right) =
\frac{\xi}{3 \pi L\left( \xi \right)} \sin \left( \alpha \pi \right)
\left\{ 
\begin{array}{lc}
\left( \frac{\tau_{k}}{\tau_{cc}} \right)^{\alpha}  & 
\text{for } \tau_{k} \ll \tau_{cc} \\ 
\left( \frac{\tau_{k}}{\tau_{cc}} \right)^{-\alpha}  & 
\text{for } \tau_{k} \gg \tau_{cc}
\end{array}
\right. .
\label{limiting014a}
\end{equation}

This result implies that the power law memory kernel (\ref{frac002})
and consequently the Cole-Cole function (\ref{limiting004}), can be 
suitable for describing the dynamics of systems in which a distribution of 
relaxation times exist, such that the fraction of polar elements of the
system participating in the processes with relaxation time $\tau_{k}$,
scales as $\tau_{k}^{\alpha}$ in the range of fast processes, while in
the range of slow processes it scales as $\tau_{k}^{-\alpha}$.
In addition, it should be noticed that the fraction of polar elements 
participating in a given relaxation process depends on the dimensionless 
parameter $\xi$ which in turn depends on the static applied field $E^{(0)}$.

\section{Conclusions}

Starting from purely thermodynamic grounds, we have formulated a very 
general framework, which is useful for describing non-Debye dielectric 
relaxation in diverse materials including solids, glass-forming liquids and 
polymer melts. The framework is based on the derivation of a non-Markovian 
Fokker-Planck like equation that incorporates memory effects for both 
rotational diffusion and diffusion of polarization and therefore provides a
very wide range of possible behaviors as analyzed through section~V. We have 
shown that fractional Fokker-Planck descriptions can be derived as particular
cases of our model when a specific form of the memory kernel is chosen. The 
formulation of the model on thermodynamic grounds simplifies its 
generalization to cases in which external temperature or pressure gradients, 
among others, are applied on the system.

The evolution equations for the polarization vector and higher order 
multipoles can be derived systematically. We analyzed the evolution equations 
for the first two multipoles by closing the corresponding hierarchy assuming 
that the higher order multipoles may be considered near to equilibrium. This 
assumption yields previously known results for the dependence of the 
susceptibilities and the relaxation times when the system evolves in the 
presence of a time dependent externally applied electric field consisting of a 
static and a harmonic dependent component. We analyzed in detail the effect 
that each mechanism considered has in the behavior of the susceptibilities and 
shown how empirical formulas can be obtained from the general formalism. In addition,
we outlined how the present approach can be generalized to consider the non-Markovian
dynamics of several polar elements and the effect of this on the total susceptibility.
In this case, our results were used to fit with recent experiments performed with
glassy materials damped with myoglobin proteins at different levels of hydration. The
agreement between theory and experiment was very good.

The analysis of the susceptibilities by means of the theory of distribution 
of times lead us to propose a physical interpretation of each relaxation 
process in terms of the fraction of polar elements that participate in the 
relaxation process for a given frequency of the external field. We found that 
this fraction of polar elements, related to a particular characteristic 
relaxation time, depends on the static externally applied field $E^{(0)}$. 
This result indicates that the application of a static field may be used to 
control the fraction of polar elements that participate in a given relaxation 
process (possibly including multistep relaxation) and thus can be useful in both 
the analysis and the design of materials 
with specific absorption windows.

\section*{Acknowledgments} 
HH acknowledges UNAM-DGAPA for financial support. 
JGMB thanks CONACYT for financial support and 
ISH thanks UNAM-DGAPA for partial financial support of Grant No. IN102609.

\appendix

\section{Auxiliar formulae}

In section~V, we have discussed how the uniaxial symmetry of the problem
reduces the matrices $M_{ij}$, $N_{ij}$ and $\chi_{ij}$  to diagonal forms
that contain two contributions, one parallel  and one perpendicular to the
static applied field. Making use of Eqs.~(\ref{perturbation011}),
(\ref{perturbation012}) and~(\ref{illustration001})-(\ref{illustration003}),
it can be shown that the explicit form of the elements $M_{\perp}$,
$M_{\parallel}$, $N_{\perp}$ and $N_{\parallel}$ are given by the expressions:
\begin{equation}
M_{\perp} \left( \vec{k}, \omega \right) = G_{1}\left( \vec{k}, \omega
\right) + \frac{3}{5} \xi^{2} \frac{\phi^{2}\left( \vec{k}, \omega \right)} {%
G_{2}\left( \vec{k}, \omega \right)},  \label{illustration005}
\end{equation}
\begin{equation}
M_{\parallel} \left( \vec{k}, \omega \right) = G_{1}\left( \vec{k}, \omega
\right) + \frac{4}{5} \xi^{2} \frac{\phi^{2}\left( \vec{k}, \omega \right)} {%
G_{2}\left( \vec{k}, \omega \right)},  \label{illustration006}
\end{equation}
\begin{equation}
N_{\perp} \left( \vec{k}, \omega \right) = 6 \chi_{0} \phi \left( \vec{k},
\omega \right) \left\{ \frac{1}{2}\left( 1 -\frac{L\left( \xi \right)} {\xi}
\right) -\frac{\phi \left( \vec{k}, \omega \right)} {G_{2}\left( \vec{k},
\omega \right)} \left[ - 1 + \left( \frac{\xi}{2} +\frac{3}{\xi} \right)
L\left( \xi \right) \right] \right\},  \label{illustration007}
\end{equation}
\begin{equation}
N_{\parallel} \left( \vec{k}, \omega \right) = 6 \chi_{0} \phi \left( \vec{k}%
, \omega \right) \left\{ \frac{L\left( \xi \right)}{\xi} +\frac{2 \phi
\left( \vec{k}, \omega \right)} {G_{2}\left( \vec{k}, \omega \right)} \left[
-1 + \frac{3}{\xi} L\left( \xi \right) \right] \right\} .
\label{illustration008}
\end{equation}
From these equations, the explicit form of the susceptibilities
Eqs.~(\ref{illustration010}) and (\ref{illustration011}) was calculated.

\section{Fitting parameters}

The parameters used for fitting the experimental data with Eq.~(\ref{Many002}) and
represented in Figs.~\ref{figure005}~(a) and (b), are given in the following table.

\begin{table}[tbh]
\begin{center}
\scriptsize{
\begin{tabular}{|c|c|c|c|c|c|c|c|c|c|c|}
\hline
Model                    & $\chi^{o}_1$ & $\chi^{o}_2$ & $\alpha_1$ & $\alpha_2$ & $\beta_1$ & $\beta_2$ & $\tau_1$ & $\tau_2$ & $\tau_{\beta_{1}}$ & $\tau_{\beta_{2}}$ \\
\hline
Fig.~\ref{figure005}~(a) & 1.368        & 0.859        & 0.715      & 0.374      & -         & -         & 0.063    & 94.470   & -                   & -                 \\
\hline
Fig.~\ref{figure005}~(b) & 4.408        & 0.978        & 0.831      & 0.013      & 0.703     & 0.541     & 0.177    & 0.673    & 0.707               & 0.421             \\
\hline
\end{tabular}
}
\end{center}
\end{table}

We have introduced the notation $\tau_q = 1/ 2D_{q}$, 
$\tau_{\beta_{q}} = 1 / k^2 \mathcal{D}_{\beta_{q}}$, for $q=1,2$.

\end{document}